\documentclass[12pt]{article}
\usepackage{epsfig}
\usepackage{amssymb}
\usepackage{amsmath}
\usepackage{amsfonts}
\usepackage{graphicx}
\usepackage{mathrsfs}
\usepackage{mathabx}
\usepackage[dvips]{color}
\usepackage{multirow}

\newcommand{\balpha}{\boldsymbol{\alpha}}
\newcommand{\bgamma}{\boldsymbol{\gamma}}
\newcommand{\bsigma}{\boldsymbol{\sigma}}

\newcommand{\bnabla}{\boldsymbol{\nabla}}

\newcommand{\R}{\mathbb{R}}
\newcommand{\C}{\mathbb{C}}

\newcommand{\fa}{\mathfrak{a}}
\newcommand{\fb}{\mathfrak{b}}
\newcommand{\fc}{\mathfrak{c}}

\newcommand{\fz}{\mathfrak{z}}

\newcommand{\fM}{\mathfrak{M}}

\newcommand{\fZ}{\mathfrak{Z}}

\newcommand{\bfa}{\mathbf{a}}

\newcommand{\bfe}{\mathbf{e}}
\newcommand{\bg}{\mathbf{g}}
\newcommand{\bh}{\mathbf{h}}
\newcommand{\bk}{\mathbf{k}}

\newcommand{\bfr}{\mathbf{r}}
\newcommand{\bt}{\mathbf{t}}
\newcommand{\bx}{\mathbf{x}}

\newcommand{\bA}{\mathbf{A}}
\newcommand{\bcA}{\boldsymbol{\cA}}
\newcommand{\bB}{\mathbf{B}}
\newcommand{\bC}{\mathbf{C}}

\newcommand{\bcE}{{\boldsymbol{\cE}}}
\newcommand{\bF}{\mathbf{F}}
\newcommand{\bG}{\mathbf{G}}
\newcommand{\bH}{\mathbf{H}}
\newcommand{\bI}{\mathbf{I}}
\newcommand{\bJ}{\mathbf{J}}

\newcommand{\bL}{\mathbf{L}}
\newcommand{\bcL}{\boldsymbol{\cL}}
\newcommand{\bM}{\mathbf{M}}
\newcommand{\bN}{\mathbf{N}}

\newcommand{\bS}{\mathbf{S}}
\newcommand{\bT}{\mathbf{T}}
\newcommand{\bU}{\mathbf{U}}

\newcommand{\bZ}{\mathbf{Z}}
\newcommand{\cA}{{\mathcal{A}}}
\newcommand{\cB}{\mathcal{B}}
\newcommand{\cC}{\mathcal{C}}

\newcommand{\cH}{\mathcal{H}}
\newcommand{\cE}{\mathcal{E}}
\newcommand{\cF}{\mathcal{F}}
\newcommand{\cG}{\mathcal{G}}
\newcommand{\bcG}{{\boldsymbol{\cG}}}
\newcommand{\cI}{\mathcal{I}}

\newcommand{\cL}{\mathcal{L}}

\newcommand{\cP}{\mathcal{P}}

\newcommand{\cT}{\mathcal{T}}
\newcommand{\cU}{\mathcal{U}}

\newcommand{\cZ}{\mathcal{Z}}
\newcommand{\be}{\begin{equation}}
\newcommand{\ee}{\end{equation}}
\newcommand{\bea}{\begin{eqnarray}}
\newcommand{\eea}{\end{eqnarray}}
\newcommand{\nn}{\nonumber}
\newcommand{\kt}{\rangle}
\newcommand{\br}{\langle}

\newcommand{\ed}{\end{document}}

\newcommand{\bi}{\begin{itemize}}
\newcommand{\ei}{\end{itemize}}

\newcommand{\bce}{\begin{center}}
\newcommand{\ece}{\end{center}}

\newcommand{\sD}{\mathscr{D}}

\newcommand{\sF}{\mathscr{F}}

\newcommand{\sT}{\mathscr{T}}

\newcommand{\RE}{{\rm Re}}
\newcommand{\IM}{{\rm Im}}

\newcommand{\bPhi}{{\boldsymbol{\Phi}}}
\newcommand{\bPi}{{\boldsymbol{\Pi}}}
\newcommand{\bcB}{{\boldsymbol{\cB}}}
\newcommand{\bcC}{{\boldsymbol{\cC}}}

\newcommand{\bfM}{{\boldsymbol{\fM}}}
\newcommand{\bcH}{{\boldsymbol{\cH}}}
\newcommand{\bcU}{{\boldsymbol{\cU}}}
\newcommand{\bfZ}{{\boldsymbol{\fZ}}}
\newcommand{\bvarepsilon}{{{\mbox{\large$\boldsymbol{\varepsilon}$}}}}
\newcommand{\bmu}{{{\mbox{\large$\boldsymbol{\mu}$}}}}

\newcommand{\bigvarepsilon}{\mbox{\large$\varepsilon$}}
\newcommand{\bigmu}{\mbox{\large$\mu$}}

\newcommand{\bzero}{{\boldsymbol{0}}}

\newcommand{\for}{{\mbox{\rm for}}}

\newcommand{\sinc}{{\rm sinc}}

\newcommand{\bup}{{\boldsymbol{\Upsilon}}}
\newcommand{\bXi}{{\boldsymbol{\Xi}}}

\newcommand{\Lpi}{{\mbox{\Large${{\pi}}$}_{\!k}}}

\begin{document}

\title{Fundamental transfer matrix for electromagnetic waves, scattering by a planar collection of point scatterers,\\ and anti-$\cP\cT$-symmetry}

%\title{Fundamental transfer matrix for electromagnetic waves propagating in a general non-homogeneous\\ and anisotropic linear medium}

\author{Farhang Loran\thanks{E-mail address: loran@iut.ac.ir}
~and Ali~Mostafazadeh\thanks{E-mail address:
amostafazadeh@ku.edu.tr} $^{, \ddagger}$\\[6pt]
$^*$Department of Physics, Isfahan University of Technology, \\ Isfahan 84156-83111, Iran\\[6pt]
$^\dagger$Departments of Mathematics and Physics, Ko\c{c} University,\\  34450 Sar{\i}yer,
Istanbul, T\"urkiye\\[6pt]
$^\ddagger$T\"{U}B$\dot{\rm I}$TAK Research Institute for Fundamental Sciences,\\ Gebze, Kocaeli 41470, T\"urkiye}

\date{ }
\maketitle

\begin{abstract}

We develop a fundamental transfer-matrix formulation of the scattering of electromagnetic (EM) waves that incorporates the contribution of the evanescent waves and applies to general stationary linear media which need not be isotropic, homogenous, or passive. Unlike the traditional transfer matrices whose definition involves slicing the medium, the fundamental transfer matrix is a linear operator acting in an infinite-dimensional function space. It is given in terms of the evolution operator for a non-unitary quantum system and has the benefit of allowing for analytic calculations. In this respect it is the only available alternative to the standard Green's-function approaches to EM scattering. We use it to offer an exact solution of the outstanding EM scattering problem for an arbitrary finite collection of possibly anisotropic nonmagnetic point scatterers lying on a plane. In particular, we provide a comprehensive treatment of doublets consisting of pairs of isotropic point scatterers and study their spectral singularities. We show that identical and $\mathcal{P}\mathcal{T}$-symmetric doublets do not admit spectral singularities and cannot function as a laser unless the real part of their permittivity equals that of vacuum. This restriction does not apply to doublets displaying anti-$\mathcal{P}\mathcal{T}$-symmetry. We determine the lasing threshold for a generic anti-$\mathcal{P}\mathcal{T}$-symmetric doublet and show that it possesses a continuous lasing spectrum. 

\vspace{2mm}

%\noindent PACS numbers: 03.65.Nk, 42.25.Bs\vspace{2mm}

%\noindent Keywords:

\end{abstract}

\section{Introduction}
\label{S1}

Transfer matrices have been used as an effective tool in the study of wave propagation in effectively one-dimensional stratified media since the 1940's  \cite{jones-1941,abeles,thompson}. They were subsequently generalized to deal with the propagation and scattering of scalar and electromagnetic (EM) waves in two and three dimensions \cite{teitler-1970,berreman-1972,pendry-1984,pendry-1990a}. These developments were guided by the basic principle of slicing the medium in which the wave propagates along a propagation/scattering axis, discretizing the transverse degrees of freedom, associating a numerical transfer matrix for each slice, and multiplying the latter according to the celebrated composition rule for transfer matrices \cite{sanchez,tjp-2020} to obtain the transfer matrix for the medium \cite{pendry-1996}. This leads to a numerical method of computing the behavior of the wave which is however plagued with instabilities arising from the multiplication of large numbers of numerical matrices. This in turn has motivated the development of various intricate variations of this approach to improve the numerical stability of the calculations \cite{tan-2003}. 

In Refs.~\cite{pra-2016,pra-2021} we follow a completely different route to define a transfer matrix for the scattering of scalar waves by an interaction potential in two and three dimensions. The result is not a numerical matrix but a linear operator acting in an infinite-dimensional function space. Similarly to its traditional numerical predecessors, this notion of transfer matrix stores the information about the scattering properties of the potential and has a build-in composition property. The latter follows from a curious relation between the transfer matrix and the evolution operator for an associated non-unitary quantum system.\footnote{A similar connection exists in one dimension  \cite{ap-2014}.} For this reason we call this approach ``dynamical formulation of the stationary scattering.'' Recently, we have proposed a similar approach for introducing a transfer matrix for the scattering of EM waves by isotropic scatterers \cite{jpa-2020a}. 

\begin{figure}
        \begin{center}
        \includegraphics[scale=.35]{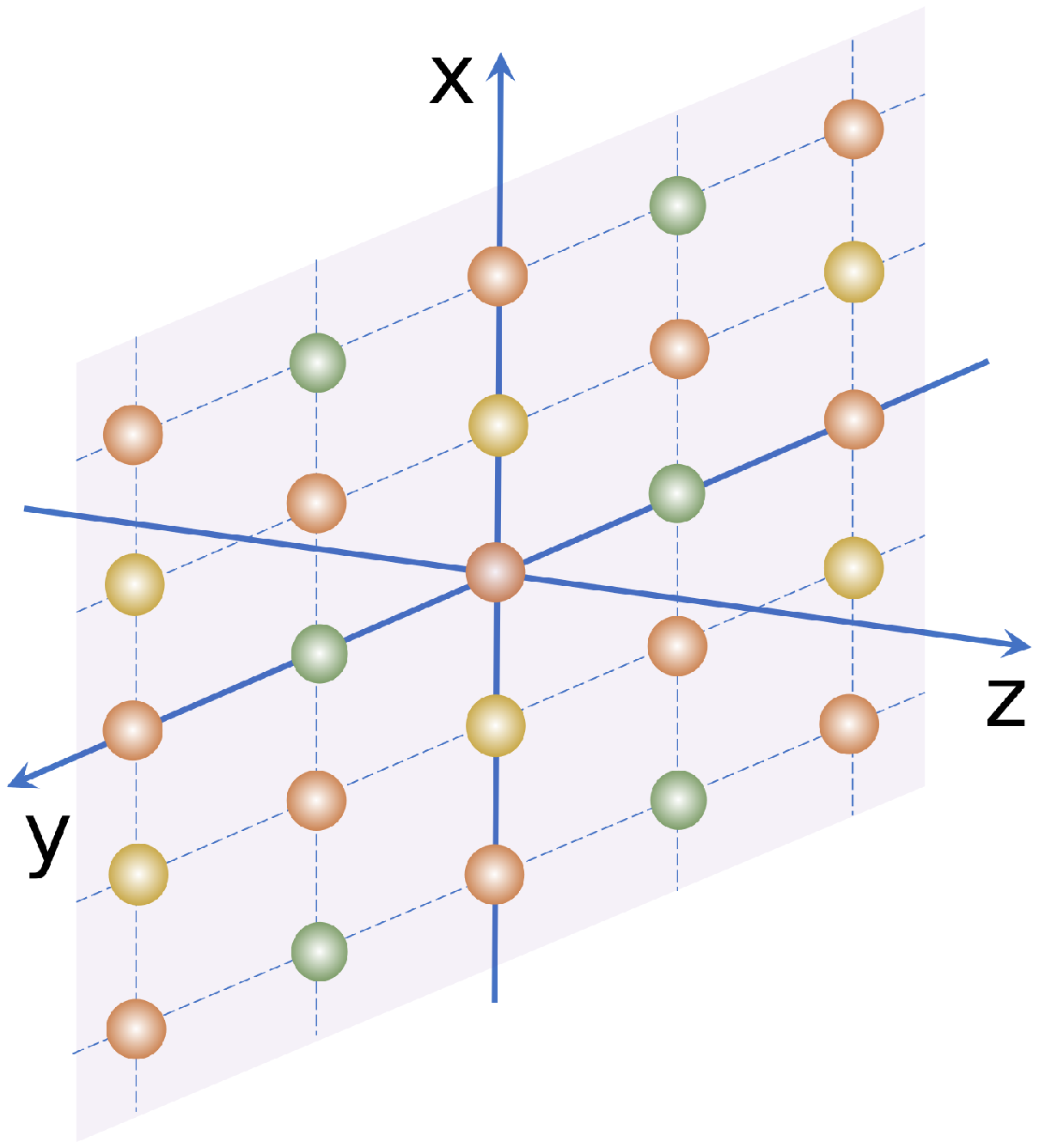}~~~~~~~~~~~~
        \includegraphics[scale=.35]{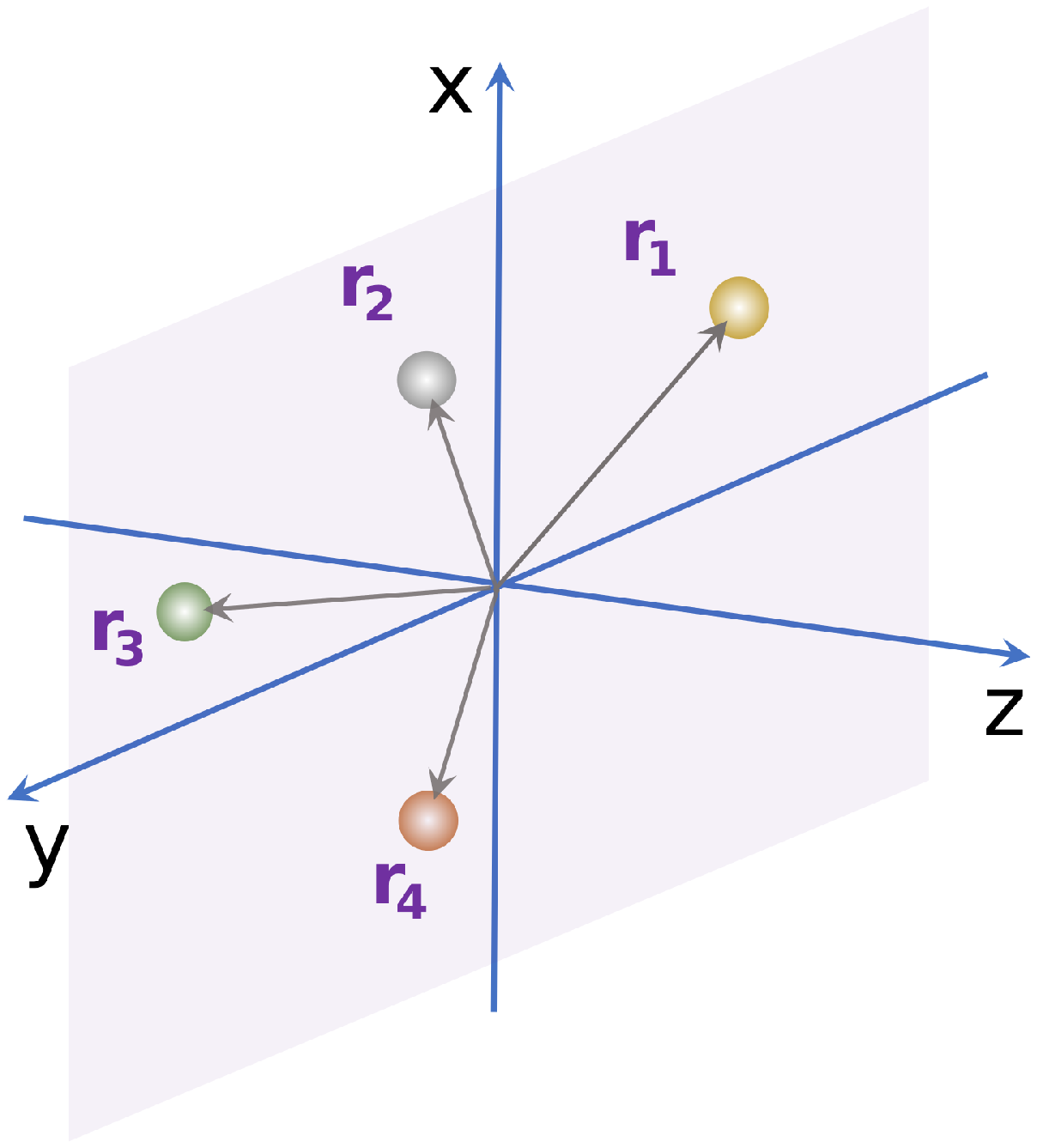}
        \caption{Schematic views of planar collections of point scatterers forming a two-dimensional regular lattice (on the left) and having arbitrary positions (on the right).}
        \label{fig1}
        \end{center}
        \end{figure}

In the first half of the present article, we develop the basic framework provided in Ref.~\cite{jpa-2020a} into a  comprehensive ``dynamical formulation of stationary scattering for EM waves.'' We achieve this by extending this framework to general anisotropic stationary linear media and introducing an EM analog of the auxiliary transfer matrix of Ref.~\cite{pra-2021} which allows us to account for the contribution of the evanescent waves, hence lifting an implicit assumption made in \cite{jpa-2020a}. An important advantage of this formulation of EM scattering over the known transfer matrix methods is that it allows for analytic calculations. In the second half of the article, we use it to obtain an exact analytic solution of the scattering problem for 
EM waves interacting with an arbitrary finite collection of nonmagnetic point scatterers that reside on a plane and can display anisotropy as well as gain and loss.\footnote{The point scatterers are confined to a plane but otherwise their positions need not have a particular symmetry. For the cases that they form a lattice, we can use this system to model certain two-dimensional optical lattices \cite{shahmoon}. {The exact analytic treatment of nonplanar configurations of point scatterers is a difficult open problem.}} If we choose a coordinate system in which the point scatterers lie on the $x$-$y$ plane, as shown in Fig.~\ref{fig1}, 
we can model them using the following permittivity and permeability tensors respectively.
	\begin{align}
	&\bvarepsilon(x,y,z)=\varepsilon_0\left[\bI+\delta(z)\sum_{a=1}^N\bfZ_a\,
	\delta(x-x_a)\delta(y-y_a)\right],
	&&\bmu\mbox{\normalsize$(x,y,z)$}=\mu_0\bI.
	\label{delta-sec1}
	\end{align}
Here $\varepsilon_0$ and $\mu_0$ are the permittivity and permeability of vacuum, $\bI$ stands for the $3\times 3$ identity matrix, $\delta(\cdot)$ is the Dirac delta function, $N$ is the number of point scatterers, $\bfZ_a$ are nonzero $3\times 3$ matrices\footnote{For technical reasons we assume that the $\fZ_{a,33}$ entry of $\bfZ_a$ is nonzero.}, and $(x_a,y_a)$ are the coordinates of the point scatterers in the $x$-$y$ plane. 

If $\bfZ_a$ are scalar multiples of the identity matrix, (\ref{delta-sec1}) represents a finite collection of isotropic point scatterers \cite{shahmoon,calla-2014}. For $N=1$ this corresponds to a single isotropic point scatterer whose scattering amplitude has been obtained in closed form in earlier studies of the subject \cite{VCL}. The standard treatment of this point scatterer leads to singularities whose removal requires the use of a highly nontrivial renormalization scheme. The application of our fundamental transfer matrix to this point scatterer turns out to avoid the unwanted singularities of its standard treatment and yields the same result \cite{jpa-2020a}. As we show in the present article, the finiteness property of this approach extends to the general case where the point scatterers need not be isotropic and their number is arbitrary. Another remarkable outcome of this approach is that the scattering amplitude for the collections of point scatterers given by (\ref{delta-sec1}) satisfies the following simple relation.
	\be
	f(\bk_{\rm i},\bk_{\rm s}) \bfe_{\rm s}=
    	\frac{k^2}{4\pi}\,\big[\hat\bfr\times(\bg\times\hat\bfr)\big].
	\label{fe=g}
	\ee
Here $f(\bk_{\rm i},\bk_{\rm s})$ stands for the scattering amplitude \cite{TKD}, $\bk_{\rm i}$ and $\bk_{\rm s}$ are respectively the incident and scattered wave vectors, $\bfe_{\rm s}$ is the polarization vector for the scattered wave, $\hat\bfr$ is the unit vector specifying the direction of $\bk_{\rm s}$, and $\bg$ is a vector lying in the $x$-$y$ plane that stores all the information about the scattering properties of the point scatterers as well as the polarization and wavevector for the incident wave.

Our approach allows for an analytic calculation of $\bg$ which simplifies considerably for doublets consisting of a pair of isotropic point scatterers. For cases where the latter are made of active optical material, the doublet may serve as a laser. Our results enable us to locate the spectral singularities \cite{prl-2009,Longhi-2010,Ramezani-2014} of such active doublets and determine their laser threshold condition \cite{pra-2011a,prsa-2012,pra-2013a,pra-2013b,hang-2016,wang-2016,jin-2018,ap-2018,konotop-2019}. A surprising outcome of this investigation is that identical and $\cP\cT$-symmetric pairs of point scatterers do not lase unless the real part of their permittivity equals that of the vacuum, a condition which makes their realization utterly difficult. This obstruction is lifted for doublets possessing anti-$\cP\cT$-symmetry. The latter is a peculiar property which has previously been investigated only in one dimension  \cite{Ge-2013,Wu-2014,Wu-2015,Peng-2016,Ge-2017,Hu-2021,He-2022}. We offer a comprehensive study of generic anti-$\cP\cT$-symmetric doublets of point scatterers determining their laser threshold condition and lasing spectrum.  

The outline of this article is as follows. In Sec.~2, we discuss a basic setup for the scattering of EM waves due to a general stationary linear medium and use it to identify the fundamental transfer matrix for these waves. Here we also describe the utility of the fundamental transfer matrix in solving EM scattering problems. In Sec.~3, we introduce an EM analog of the auxiliary transfer matrix of Ref.~\cite{pra-2021}, reveal its relationship to the fundamental transfer matrix, and derive its Dyson series expansion.  In Sec.~4, we employ our general results to obtain an exact solution of the scattering problem for planar collections of point scatterer. Here {we derive Eq.~(\ref{fe=g}), give the explicit form of the vector $\bg$ in terms of the physical parameters of the system,} offer a detailed treatment of doublets of point scatterers, and examine their spectral singularities. In Sec.~5, we confine our attention to the study of anti-$\cP\cT$-symmetric pairs of isotropic point scatterers and determine their laser threshold condition and lasing spectrum. Sec.~6 presents our concluding remarks.

Throughout this article, we use the following basic notations and conventions.
	\begin{itemize}
	\item[-] $\C^{m\times n}$ denotes the set of complex $m\times n$ matrices. In particular, $\C^{m\times 1}$ consists of column vectors with $m$ components.
	\item[-] $\sF^{m}$ denotes the set of (generalized) functions that map $\R^2$ to $\C^{m\times 1}$, i.e., its elements are $m$-component functions; if $\bF\in\sF^m$ and $\vec r\in\R^2$, there are functions $F_1,F_2,\cdots,F_m\in\sF^1$ such that
		\[\bF(\vec r)=\left[\begin{array}{c}
		F_1(\vec r)\\
		F_2(\vec r)\\
		\vdots\\
		F_m(\vec r)\end{array}\right].\]
	\item[-] $\bzero$ and $\bI$ label the zero and identity matrices of appropriate size, and $\widehat\bzero$ and $\widehat\bI$ mark the zero and identity operators acting in the relevant function spaces, respectively. 
	\item[-] We adopt a Cartesian coordinate system $\{(x,y,z)\}$ such that the source of the incident wave and the detectors measuring the scattered wave lie on the planes $z=\pm\infty$. See Fig.~\ref{fig2}
	\begin{figure}
        \begin{center}
        \includegraphics[scale=.35]{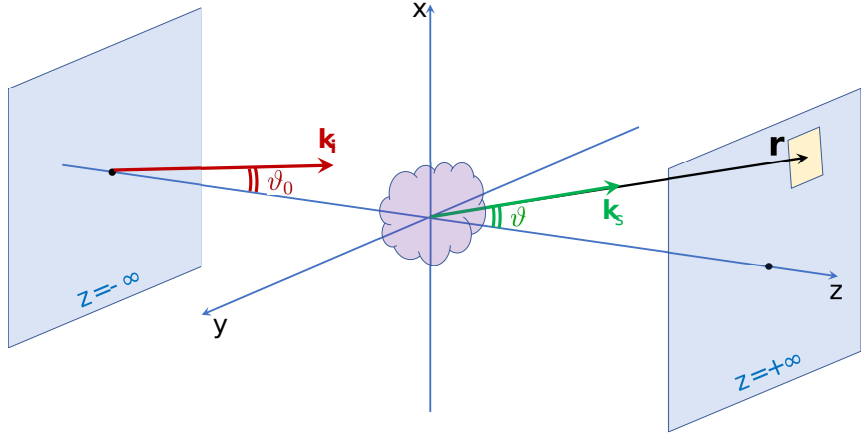}%\vspace{-5cm}
        \caption{Schematic view of the scattering setup for a left-incident plane wave: The source of the incident wave is located at $z=-\infty$. The purple region contains the scattering medium. $\bfr$ is the position of a detector placed at $z=+\infty$. $\bk_{\rm i}$ and $\bk_{\rm s}$ are respectively the incident and scattered wave vectors. $\vartheta_0$ and $\vartheta$ are the angles they make with the positive $z$-axis.}
        \label{fig2}
        \end{center}
        \end{figure}
	\item[-] $\bfe_x,\bfe_y$, and $\bfe_z$ denote the unit vectors pointing along the $x$-, $y$-, and $z$-axes, respectively.
	\item[-] If $\bfa=(a_x,a_y,a_z)\in\R^3$, we use $\vec a$ to denote the projection of $\bfa$ onto the $x$-$y$ plane;	i.e., $\vec a:=a_x\bfe_x+a_y\bfe_y$. We often identify $\vec a$ with $(a_x,a_y)\in\R^2$, and use the hybrid notation, $\bfa=(\vec a,a_z)$. In particular, we write the position vector, $\bfr:=(x,y,z)$, also in the form $(\vec r,z)$. 
	\end{itemize}

\section{Scattering of EM waves and fundamental transfer matrix}

\subsection{Scattering amplitude and differential cross section}

Consider the propagation of time-harmonic EM waves in a stationary linear dielectric medium that does not include any free charges or currents. Let $\bvarepsilon(\bfr)$ and $\bmu(\bfr)$ be the permittivity and permeability tensors for the medium, respectively. Then the electric and magnetic fields associated with this wave take the form, $e^{-i\omega t}\bcE(\bfr)/\sqrt\varepsilon_0$ and $e^{-i\omega t}\bcH(\bfr)/\sqrt\mu_0$, where $\omega$ is the angular frequency of the wave, Maxwell equations imply 
	\begin{align}
	&\bnabla\cdot(\hat\bvarepsilon\,\bcE)=0,
	&&\bnabla\cdot(\hat\bmu\,\bcH)=0,
	\label{mx1}\\
	&\bnabla\times\bcE=ik\hat\bmu\,\bcH,
	&&\bnabla\times\bcH=-ik\hat\bvarepsilon\,\bcE,
	\label{mx2}
	\end{align}
$\hat\bvarepsilon$$(\bfr):=\varepsilon_0^{-1}\bvarepsilon(\bfr)$ and $\hat\bmu$$(\bfr):=\mu_0^{-1}\bmu(\bfr)$ are the relative permittivity and permeability tensors, and $k:=\omega/c$ is the wavenumber.

Suppose that for $r:=|\bfr| \to\infty$, ${\hat\bvarepsilon}$$(\bfr)- 
\bI$ and $\hat\bmu$$(\bfr)-\bI$ tend to $\bzero$ at such a rate that (\ref{mx1}) and (\ref{mx2}) admit solutions fulfilling the asymptotic boundary condition:
    \be
    \bcE(\bfr)= \cE_0 \left[ e^{i\bk_{\rm i}\cdot\bfr} \bfe_{\rm i}+\frac{e^{ikr}}{r}\,f(\bk_{\rm s},\bk_{\rm i})\,\bfe_{\rm s}\right]~~
    {\rm for}~~r\to\infty,
    \label{asym-BC}
    \ee
where $\cE_0$ is a constant, $\bk_{\rm i}$ and $\bk_{\rm s}:=k\bfr/r=k\hat\bfr$ are respectively the wavevectors for the incident and scattered waves, $\bfe_{\rm i}$ and $\bfe_{\rm s}$ are the polarization vectors for the incident and scattered waves, and $f(\bk_{\rm s},\bk_{\rm i})$ is the scattering amplitude. The latter determines the differential cross section\footnote{By definition, $\sigma_d(\bk_{\rm s},\bk_{\rm i}):= r^2 |\br\bS_{\rm s}\kt|/|\br\bS_{\rm i}\kt|$, where $\br \bS_{\rm i}\kt$ and $\br \bS_{\rm s}\kt$ are the time-averaged Poynting vectors for the incident and scattered waves \cite{TKD}.} according to
    \be
    \sigma_d(\bk_{\rm s},\bk_{\rm i})=\left|f(\bk_{\rm s},\bk_{\rm i})\right|^2.
    \label{diff-cross-sec}
    \ee
The first and second terms in the square bracket in (\ref{asym-BC}) respectively correspond to the incident and scattered waves;
    \begin{align}
    &\bcE_{\rm i}(\bfr):=\cE_0 e^{i\bk_{\rm i}\cdot\bfr} \bfe_{\rm i},
    &&\bcE_{\rm s}(\bfr):= \frac{\cE_0\,e^{ikr}}{r}\,f(\bk_{\rm s},\bk_{\rm i})\,\bfe_{\rm s}.
    \label{Ei-Es}
    \end{align}
Their wave and polarization vectors satisfy:
    \begin{align*}
    &|\bk_{\rm i}|=|\bk_{\rm s}|=k,
    &&\bfe_{\rm i}\cdot\bk_{\rm i}=0,
    &&\bfe_{\rm s}\cdot\bk_{\rm s}=k\,\bfe_{\rm s}\cdot\hat\bfr=0.
    \end{align*}
By solving the scattering problem for the medium, we mean the determination of the scattered wave, alternatively $f(\bk_{\rm s},\bk_{\rm i})\,\bfe_{\rm s}$, which is a function of the wavenumber and polarization of the incident wave, $k$ and $\bfe_{\rm i}$, and the directions of the incident and scattered wavevectors, $\hat\bk_{\rm i}=\bk_{\rm i}/k$ and $\hat\bk_{\rm s}=\hat\bfr$.

We also recall that the magnetic fields for the incident/scattered waves are given by $\bcH_{\rm i,s}=c^{-1}\hat\bk_{\rm i,s}\times\bcE_{\rm i,s}$. In particular, they are scalar multiples of the following unit vectors, 
	\begin{align}
	&\bh_{\rm i}=\hat\bk_{\rm i}\times\bfe_{\rm i},
	&&\bh_{\rm s}=\hat\bk_{\rm s}\times\bfe_{\rm s}=
	\hat\bfr\times\bfe_{\rm s}.
	\label{bhs-def}
	\end{align}

\subsection{Effective Schr\"odinger equation for time-harmonic EM waves}

Consider the dynamical Maxwell equations (\ref{mx2}), supposing that the $\hat\varepsilon_{33}$ and $\hat\mu_{33}$ entries of the relative permittivity and relative permeability tensors do not vanish identically, we can use these equations to express $\cE_z$ and $\cH_z$ in terms of the $\cE_x,\cE_y,\cH_x$ and $\cH_y$, \cite{teitler-1970,berreman-1972}. Specifically, we have
	\begin{align}
	&\cE_z=\hat\varepsilon_{33}^{-1}\left[
	-\hat\varepsilon_{31}\cE_x-\hat\varepsilon_{32}\cE_y+
	\frac{i}{k}\left(\partial_x\cH_y-\partial_y\cH_x\right)\right],
	\label{Ez=}\\
	&\cH_z=\hat\mu_{33}^{-1}\left[
	-\frac{i}{k}\left(\partial_x\cE_y-\partial_y\cE_x\right)
	-\hat\mu_{31}\cH_x-\hat\mu_{32}\cH_y\right].
	\label{Hz=}
	\end{align}
In view of these relations, we can reduce (\ref{mx2}) to a system of four first-order differential equations for $\cE_x,\cE_y,\cH_x$ and $\cH_y$. It is not difficult to see that this is equivalent to the time-dependent Schr\"odinger equation,
	\be
	i\partial_z\bPhi(x,y,z)=\widehat\bH\,\bPhi(x,y,z),
	\label{TD-sch-eq}
	\ee
for the $4$-component field \cite{berreman-1972,jpa-2020a},
	\be
	\bPhi :=\left[\begin{array}{c}
	\cE_x \\ \cE_y \\ 
	\cH_x  \\ \cH_y 	
	\end{array}\right]=\left[\begin{array}{c}
	\vec\cE \\ \vec\cH \end{array}\right],
	\label{4-com-field}
	\ee
where $\vec\cE :=\left[\begin{array}{c}
	\cE_x \\ \cE_y \end{array}\right]$, 
	$\vec\cH :=\left[\begin{array}{c}
	\cH_x \\ \cH_y \end{array}\right]$,
$z$ plays the role of `time,' and $\widehat\bH$ is a `time-dependent' $4\times 4$ matrix Hamiltonian with operator entries. 

To derive an explicit expression for $\widehat\bH$, first we introduce:
	\begin{align}
	&\vec\bigvarepsilon_\ell :=\left[\begin{array}{c}
	\hat\varepsilon_{\ell 1} \\
	\hat\varepsilon_{\ell 2} \end{array}\right],
	\quad\quad\quad\quad\quad
	\vec\bigmu_\ell :=\left[\begin{array}{c}
	\hat\mu_{\ell 1} \\
	\hat\mu_{\ell 2} \end{array}\right],
	\quad\quad\quad\quad\quad
	\vec\partial:=\left[\begin{array}{c}
	\partial_x \\
	\partial_y \end{array}\right],
	\label{new1}\\
	&\widehat\bL_{\cE}:=\hat\varepsilon_{33}^{-1}\left[
	\begin{array}{c}
	-\vec\bigvarepsilon_3\\
	k^{-1}\bsigma_2\vec\partial\end{array}\right],
	\quad\quad\quad\quad\quad\quad\quad
	\widehat\bL_{\cH}:=-\hat\mu_{33}^{-1}\left[
	\begin{array}{c}
	k^{-1}\bsigma_2\vec\partial\\
	\vec\bigmu_3
	\end{array}\right],
	\label{LE-LH=}
	\end{align}
where $\ell\in\{1,2,3\}$, and express (\ref{Ez=}) and (\ref{Hz=}) in the form,
	\begin{align}
	&\cE_z=-\hat\varepsilon_{33}^{-1}\Big(\vec\bigvarepsilon_3^{\;T}\vec\cE
	+k^{-1}\vec\partial^{\;T}\bsigma_2\vec\cH\Big)=\widehat\bL_{\cE}^T\,\bPhi,
	&&\cH_z=\hat\mu_{33}^{-1}\Big(
	k^{-1}\vec\partial^{\;T}\bsigma_2\,\vec\cE
	-\vec\bigmu_3^{\;T}\vec\cH\Big)=\widehat\bL_{\cH}^T\,\bPhi,
	\label{EzHz=}
	\end{align}
where a superscript `T' marks the transpose of a matrix, and $\bsigma_\ell$ are the Pauli matrices,
	\begin{align*}
	&\bsigma_1:=\left[\begin{array}{cc}
	0 & 1\\
	1 & 0\end{array}\right],
	&&\bsigma_2:=\left[\begin{array}{cc}
	0 & -i\\
	i & 0\end{array}\right],
	&&\bsigma_3:=\left[\begin{array}{cc}
	1 & 0\\
	0 & -1\end{array}\right].
	\end{align*}

Next, we solve (\ref{mx2}) for $\partial_z\cE_x, \partial_z\cE_y, \partial_z\cH_x$, and $\partial_z\cH_y$, and use (\ref{EzHz=}) to establish
	\begin{align}
	&i\partial_z\vec\cE=k\left(\bJ_\cH\vec\cH+
	\vec J_\cH\widehat\bL_{\cH}^T \bPhi\right)+
	i\vec\partial\;\widehat\bL_{\cE}^T \bPhi,
	\label{eq13}\\
	&i\partial_z\vec\cH=-k\left(\bJ_\cE\vec\cE+
	\vec J_\cE\widehat\bL_{\cE}^T \bPhi\right)+
	i\vec\partial\;\widehat\bL_{\cH}^T \bPhi,
	\label{eq14}
	\end{align}
where
	\begin{align}
	&\vec J_\cE:=\Bigg[\begin{array}{c}
	-\hat\varepsilon_{23}\\
	\hat\varepsilon_{13}\end{array}\Bigg],
	&&\bJ_\cE:=\Bigg[\begin{array}{c}
	-\vec\bigvarepsilon_2^{\,T}\\
	\vec\bigvarepsilon_1^{\,T}\end{array}\Bigg]=
	\Bigg[\begin{array}{cc}
	-\hat\varepsilon_{21} & -\hat\varepsilon_{22}\\
	\hat\varepsilon_{11} & \hat\varepsilon_{12}
	\end{array}\Bigg],
	\label{JEs=}\\
	&\vec J_\cH:=\Bigg[\begin{array}{c}
	-\hat\mu_{23}\\
	\hat\mu_{13}\end{array}\Bigg],
	&&\bJ_\cH:=\Bigg[\begin{array}{c}
	-\vec\bigmu_2^{\,T}\\
	\vec\bigmu_1^{\,T}\end{array}\Bigg]=
	\Bigg[\begin{array}{cc}
	-\hat\mu_{21} & -\hat\mu_{22}\\
	\hat\mu_{11} & \hat\mu_{12}
	\end{array}\Bigg].
	\label{JHs=}
	\end{align}
Eqs.~(\ref{4-com-field}), (\ref{eq13}), and (\ref{eq14}) imply the Schr\"odinger equation (\ref{TD-sch-eq}) with $\widehat\bH$ given by
	\be
	\widehat\bH:=\left[\begin{array}{cc}
	\widehat\bH_{11} & \widehat\bH_{12} \\[6pt]
	\widehat\bH_{21}  & \widehat\bH_{22}\end{array}\right],
	\label{H-def}
	\ee
	\vspace{-18pt}
	\begin{align}
	&{\widehat\bH}_{11}:=-i{\vec\partial}\left(\frac{\vec\bigvarepsilon_3^{\,T}}{\hat\varepsilon_{33}}\right)+
	\frac{1}{\hat\mu_{33}}\vec J_\cH\vec\partial^{\,T}\bsigma_2,
    	&&{\widehat\bH}_{12}:=-\frac{i}{k}{\vec\partial}\left(\frac{1}{\hat\varepsilon_{33}}
	\vec\partial^{\,T}\bsigma_2\right) +k(\bJ_\cH-\tilde\bJ_\cH),
	\label{H1}\\
    	&{\widehat\bH}_{21}:=\frac{i}{k}{\vec\partial}\left(\frac{1}{\hat\mu_{33}}
	\vec\partial^{\,T}\bsigma_2\right)+ k(\tilde\bJ_\cE-\bJ_\cE),
    	&&{\widehat\bH}_{22}:=-i{\vec\partial}\left(\frac{\vec\bigmu_3^{\,T}}{\hat\mu_{33}}\right)+
	\frac{1}{\hat\varepsilon_{33}}\vec J_\cE\vec\partial^{\,T}\bsigma_2,
	\label{H2}\\
    	&\tilde\bJ_\cE:=\frac{1}{\hat\varepsilon_{33}}\vec J_\cE\vec\bigvarepsilon_3^{\,T},
	&&\tilde\bJ_\cH:=\frac{1}{\hat\mu_{33}}\vec J_\cH\vec\bigmu_3^{\,T}.
	\label{tJs=}
	\end{align}

\subsection{Fundamental transfer matrix for EM waves}
	
For $r\to\infty$, $\hat\bvarepsilon\mbox{\normalsize$(\bfr)\to\bI$}$, $\hat\bmu\mbox{\normalsize$(\bfr)\to \bI$}$, 
	\begin{align}
	&\widehat\bH\to \widehat\bH_0:=\left[\begin{array}{cc}
    	\bzero & {\widehat{\bL}}_0\\
    	- {\widehat{\bL}}_0 & \bzero\end{array}\right],
	\label{H-zero}
    	\end{align}
where
	\be
	{\widehat{\bL}}_0:=-ik^{-1}(\vec\partial\vec\partial^{\,T}+k^2\bI)\bsigma_2=
	k^{-1}\left[\begin{array}{cc}
	\partial_x\partial_y&-\partial_x^2-k^2\\
	\partial_y^2+k^2 & -\partial_x\partial_y\end{array}\right],
	\label{L-zero=}
	\ee	
and solutions $\bPhi$ of (\ref{TD-sch-eq}) tend to those of 
	\be
	i\partial_z\bPhi_0(x,y,z)=\widehat\bH_0\bPhi_0(x,y,z).
	\label{Sch-eq}
	\ee 
Because $\widehat\bH_0^2=(\partial_x^2+\partial_y^2+k^2)\bI$, for each choice of $z$, $\bPhi_0(\cdot,\cdot,z)$ solves the Helmholtz equation in two dimensions \cite{jpa-2020a}. Therefore,  
	\be
	\bPhi_0(x,y,z)=\bPhi_0(\vec r,z)=\frac{1}{4\pi^2} \int_{\R^2} d^2\vec p\: e^{i\vec p\cdot\vec r}
         \left[ \bcA(\vec p) e^{i\varpi(\vec p)z}+ \bcB(\vec p)e^{-i\varpi(\vec p)z}\right],
         \label{Phi-zero}
         \ee
where $\bcA,\bcB\in\sF^4$ are 4-component coefficient functions, and
	\be
	\varpi(\vec p):=\left\{\begin{array}{ccc}
	\sqrt{k^2-\vec p^{\:2}} & \for & |\vec p|<k,\\[3pt]
	i\sqrt{\vec p^{\:2}-k^2} & \for & |\vec p|\geq k.
	\end{array}\right.
	\ee

Performing the Fourier transform of both sides of (\ref{Phi-zero}) with respect to $x$ and $y$, we find 
	\be
	i\partial_z\tilde\bPhi_0(\vec p,z)=\tilde\bH_0(\vec p\,)\tilde\bPhi_0(\vec p,z),
	\label{sch-eq-zero-FT}
	\ee
where $\tilde\bPhi_0(\vec p,z)$ is the Fourier transform of $\bPhi(\vec r,z):=\bPhi(x,y,z)$ with respect to $\vec r=(x,y)$, i.e.,
	\be
	\tilde\bPhi_0(\vec p,z):=\int_{\R^2}d\vec r^{\:2}\: e^{-i\vec p\cdot\vec r}\bPhi_0(\vec r,z)=
	\bcA(\vec p\,) e^{i\varpi(\vec p)z}+ \bcB(\vec p\,)e^{-i\varpi(\vec p)z},
	\label{Phi-tilde}
	\ee 
and
	\begin{align}
	&\tilde{\bH}_0(\vec p):=\left[\begin{array}{cc}
    	\bzero & \tilde{\bL}_0(\vec p)\\
    	- \tilde{\bL}_0(\vec p) & \bzero\end{array}\right],
    	&& \tilde{\bL}_0(\vec p):=\frac{1}{k}\left[\begin{array}{cc}
    	-p_xp_y & p_x^2-k^2\\
   	-p_y^2+k^2 & p_xp_y\end{array}\right].
    	\label{t-H-zero}
    	\end{align}   
Because $\tilde{\bH}_0(\vec p)$ does not depend on $z$, we can express the general solution of (\ref{sch-eq-zero-FT}) in the form,
	\be
	\tilde\bPhi_0(\vec p,z)=e^{-iz\tilde\bH_0(\vec p)}\bcC(\vec p),
	\label{Phi-zero=}
	\ee
where 
	\be
	\bcC(\vec p):=\tilde\bPhi_0(\vec p,0)=\bcA(\vec p)+\bcB(\vec p).
	\label{cC=cA+cB}
	\ee 
Furthermore, we can use (\ref{sch-eq-zero-FT}) and (\ref{Phi-tilde}) to identify $\bcA(\vec p)$ and $\bcB(\vec p)$ with eigenvectors of $\tilde\bH_0(\vec p)$ with eigenvalues $-\varpi(\vec p)$ and $\varpi(\vec p)$, respectively \cite{jpa-2020a};
	\begin{align}
	&\tilde\bH_0(\vec p)\bcA(\vec p)=-\varpi(\vec p)\bcA(\vec p),
	&&\tilde\bH_0(\vec p)\bcB(\vec p)=\varpi(\vec p)\bcB(\vec p).
	\label{eg-va}
	\end{align}
For $|\vec p|\neq k$, we can introduce the projection matrices,
	\be
	\bPi_j(\vec p):=\frac{1}{2}\left[\bI
    	+\frac{(-1)^j}{\varpi(\vec p)}\,\tilde{\bH}_0(\vec p)\right]
	=\frac{1}{2\varpi(\vec p)}\left[
	\begin{array}{cc}
	\varpi(\vec p)&(-1)^j\tilde\bL_0(\vec p)\\
	(-1)^{j+1}\tilde\bL_0(\vec p)&\varpi(\vec p)
	\end{array}\right],~~~~~j\in\{1,2\},
    	\label{proj}
    	\ee
onto the eigenspaces associated with the eigenvalues $(-1)^j\varpi(\vec p)$ of $\tilde\bH_0(\vec p)$, and show that 
	\begin{align}
	&\bPi_1(\vec p)+\bPi_2(\vec p)=\bI,
	&&\bPi_i(\vec p)\bPi_j(\vec p)=\delta_{ij}\bPi_j(\vec p),
	\label{ortho-proj}\\
	&\bcA(\vec p)=\bPi_1(\vec p)\bcC(\vec p),
	&&\bcB(\vec p)=\bPi_2(\vec p)\bcC(\vec p),
	\label{id-1}
	\end{align}
where $\delta_{ij}$ is the Kronecker delta symbol. According to (\ref{id-1}), $\bcC$ determines $\bcA$ and $\bcB$ uniquely.

For $|\vec p|<k$, $\varpi(\vec p)$ takes positive real values and the corresponding Fourier modes on the right-hand side of (\ref{Phi-zero}) generate the right- and left-going plane-wave solutions of (\ref{Sch-eq}). Every solution of this equation is the sum of such an oscillating plane-wave solution $\bPhi_{0,{\rm os}}$ and an evanescent wave solution $\bPhi_{0, {\rm ev}}$; $\bPhi_0=\bPhi_{0,{\rm os}}+\bPhi_{0, {\rm ev}}$ where
	\bea
	\bPhi_{0,{\rm os}}(\vec r,z)&=&\frac{1}{4\pi^2} \int_{\sD_k} d^2\vec p\: e^{i\vec p\cdot\vec r}
         \left[ \bA(\vec p) e^{i\varpi(\vec p)z}+ \bB(\vec p)e^{-i\varpi(\vec p)z}\right],
         \label{os}\\
         \bPhi_{0,{\rm ev}}(\vec r,z)&=&\frac{1}{4\pi^2} \int_{\R^2\setminus\sD_k} d^2\vec p\: e^{i\vec p\cdot\vec r}
         \left[ \check\bA(\vec p) e^{i\varpi(\vec p)z}+ \check\bB(\vec p)e^{-i\varpi(\vec p)z}\right],
         \label{ev}
         \eea
where $\sD_k:=\big\{\vec p\in\R^2~\big|~|\vec p|<k\big\}$,
	\begin{align}
	&\bA:=\widehat\Lpi\bcA,
	&&\bB:=\widehat\Lpi\bcB,
	&&\check\bA:=(\widehat\bI-\widehat\Lpi)\bcA,
	&&\check\bB:=(\widehat\bI-\widehat\Lpi)\bcB,
	\label{AB}
	\end{align}
and $\widehat\Lpi:\sF^4\to\sF^4$ is the projection operator, 
	\be
	\big(\widehat\Lpi\bF\big)(\vec p):=\left\{\begin{array}{ccc}
	\bF(\vec p)&\for&|\vec p|<k,\\
	\bzero&\for&|\vec p|\geq k,\end{array}\right.
	\label{project}
	\ee
that maps $\sF^4$ onto 
	\[\sF^4_k:=\left\{\bF\in\sF^4\big|\bF(\vec p)=\bzero~\for~|\vec p|\geq 0\right\}.\]
 
Next, consider the case that the scattering medium lies between a pair of planes $z=a_\pm$, where $a_-<a_+$, i.e., for $z\notin (a_-,a_+)$, $\hat\bvarepsilon\mbox{\normalsize$(\bfr)=\bI$}$ and $\hat\bmu\mbox{\normalsize$(\bfr)=\bI$}$.\footnote{This is always the case for a finite-size scatterer. We can recover the general case by letting $a_\pm\to\pm\infty$.} The EM field configurations that do not grow exponentially as $z\to\pm\infty$ correspond to bounded solutions of the Schr\"odinger equation (\ref{TD-sch-eq}). These have the form, 
	\be
	\bPhi=\bPhi_{\rm os}+\bPhi_{\rm ev},
	\label{decompose}
	\ee 
where
	\bea
	\bPhi(\vec r,z)&=&\frac{1}{4\pi^2} \int_{\R^2} d^2\vec p\: e^{i\vec p\cdot\vec r}\times
	\left\{\begin{array}{ccc}
	\bcA_-(\vec p\,) e^{i\varpi(\vec p)z}+ \bcB_-(\vec p\,)e^{-i\varpi(\vec p)z} &\for & z<a_-,\\[3pt]
	\bcA_+(\vec p\,) e^{i\varpi(\vec p)z}+ \bcB_+(\vec p\,)e^{-i\varpi(\vec p)z} &\for & z>a_+,\end{array}\right.
    	\label{asym-EM-2}\\[6pt]
	\bPhi_{\rm os}(\vec r,z)&=&\frac{1}{4\pi^2} \int_{\sD_k} d^2\vec p\: e^{i\vec p\cdot\vec r}
	\times
	\left\{\begin{array}{ccc}
    	\bA_-(\vec p\,) e^{i\varpi(\vec p)z}+ \bB_-(\vec p\,)e^{-i\varpi(\vec p)z}&\for & z<a_-,\\[3pt]
	\bA_+(\vec p\,) e^{i\varpi(\vec p)z}+ \bB_+(\vec p\,)e^{-i\varpi(\vec p)z}&\for & z>a_+,\end{array}\right.	
    	\label{asym-EM-2-os}\\[6pt]
    	\bPhi_{\rm ev}(\vec r,z)&=&\frac{1}{4\pi^2} \int_{\R^2\setminus\sD_k}\!\!\! d^2\vec p\: 
	e^{i\vec p\cdot\vec r}\times
	\left\{\begin{array}{ccc}
	\check\bB_-(\vec p\,)\,e^{|\varpi(\vec p)|z} &\for &z<a_-,\\[3pt]
	\check\bA_+(\vec p\,)\,e^{-|\varpi(\vec p)|z} &\for &z>a_+,\end{array}\right.
    	\label{asym-EM-2-ev}
    	\eea
$\bcA_\pm,\bcB_\pm\in\sF^4$, and 
	\begin{align}
	&\bA_\pm:=\widehat\Lpi\bcA_\pm,
	&&\bB_\pm:=\widehat\Lpi\bcB_\pm, 
	&&\check\bA_+:=\bcA_+-\bA_+,
	&&\check\bB_-:=\bcB_--\bB_-.
	\end{align}
Let us also introduce
	\begin{align}
	&\bcC_\pm:=\bcA_\pm+\bcB_\pm,
	&&\bC_\pm:=\widehat\Lpi\,\bcC_\pm=\bA_\pm+\bB_\pm.
	\label{cbC-bC}
	\end{align}
Then, setting $\bcC=\bcC_\pm$ in (\ref{id-1}), we find
	\begin{align}
	&\bA_\pm=\widehat\bPi_1\bC_\pm,
	&&\bB_\pm=\widehat\bPi_2\bC_\pm,
	\label{project-pm}
	\end{align}
where $\widehat\bPi_j:\sF_k^4\to\sF_k^4$ is the projection operators defined by
	\be
	(\widehat\bPi_j\bF)(\vec p):=\bPi_j(\vec p)\bF(\vec p),
	\label{proj12-def}
	\ee
$j\in\{1,2\}$, and $\vec p\in\sD_k$. Note that, in view of (\ref{ortho-proj}), (\ref{project}), (\ref{project-pm}), and (\ref{proj12-def}),
	\begin{align}
	&\widehat\bPi_1\bA_\pm=\bA_\pm,
	&&\widehat\bPi_2\bB_\pm=\bB_\pm,
	&&\widehat\bPi_2\bA_\pm=\widehat\bPi_1\bB_\pm=\widehat\bzero,
	&&[\widehat\Lpi,\widehat\bPi_j]=\widehat\bzero.
	\label{proj-Ids}
	\end{align}

For $z\to\pm\infty$, $\bPhi_{\rm ev}(\vec r,z)\to\bzero$. Therefore,
	\be
	\bPhi(\vec r,z)\to\bPhi_{\rm os}(\vec r,z)~~~~\for~~~~z\to\pm\infty,
	\label{asymp-11}
	\ee
and the asymptotic behavior of $\bPhi(\vec r,z)$ is determined by $\bA_\pm$ and $\bB_\pm$, or alternatively by $\bC_\pm$. This provides our basic motivation for identifying the fundamental transfer matrix for EM waves with a linear operator $\widehat\bM:\sF^4\to\sF^4$ that satisfies,
	\be
	\bC_+=\widehat\bM\,\bC_-.
	\label{M-def}
	\ee

\subsection{Solution of the scattering problem}	
	
Similarly to the traditional numerical transfer matrices \cite{jones-1941,abeles,thompson,teitler-1970,berreman-1972,pendry-1984,pendry-1990a,sanchez,tjp-2020,pendry-1996} and the fundamental transfer matrix for scalar waves \cite{pra-2021}, we can use $\widehat\bM$ to determine the scattering amplitude. This offers an alternative procedure for solving EM scattering problems whose details we have reported in Ref.~\cite{jpa-2020a}. Here we provide a brief summary of this procedure and elaborate on its utility in obtaining a series expansion for the scattering amplitude. 

For the cases where the source of the incident wave is located at $z=-\infty$, as depicted in Fig.~\ref{fig2}, we proceed as follows. 
\begin{enumerate}
\item Introduce
	\begin{align}
	&\vec\bfe_{\rm i,s}:=\left[\begin{array}{c}
	\bfe_{\rm i,s}\cdot\bfe_x\\
	\bfe_{\rm i,s}\cdot\bfe_y
	\end{array}\right],
	&&\vec\bh_{\rm i,s}:=\left[\begin{array}{c}
	\bh_{\rm i,s}\cdot\bfe_x\\
	\bh_{\rm i,s}\cdot\bfe_y
	\end{array}\right],
	&&\bup_{\rm i,s}:=\left[\begin{array}{c}
	\vec\bfe_{\rm i,s}\\
	\vec\bh_{\rm i,s}\end{array}\right],
	\label{incident}
	\end{align}
where we recall that $\bfe_{\rm i}$ and $\bfe_{\rm s}$ are the polarization vectors for the incident and scattered waves, $\bh_{\rm i}$ and $\bh_{\rm s}$ are the unit vectors (\ref{bhs-def}), which signify the direction of the magnetic field for these waves, and $\vec\bfe_{\rm i}$, $\vec\bh_{\rm i}$, $\vec\bfe_{\rm s}$, and $\vec\bh_{\rm s}$ represent the projections of $\bfe_{\rm i}$, $\bh_{\rm s}$, $\bfe_{\rm s}$ and $\bh_{\rm s}$ onto the $x$-$y$ plane, respectively. It is easy to see from (\ref{bhs-def}) that
	\begin{align}
	&\vec\bh_{\rm i}=\left[\begin{array}{c}
	(-\cos{\vartheta_0}\: \bfe_y+
    	\sin\vartheta_0\sin\varphi_0\:\bfe_z)\cdot\bfe_{\rm i}\\
    	(\cos\vartheta_0\: \bfe_x
    	-\sin\vartheta_0\cos\varphi_0\:\bfe_z)\cdot\bfe_{\rm i}
	\end{array}\right],
	\label{hi=}\\
	&\vec\bh_{\rm s}=\left[\begin{array}{c}	
	(-\cos{\vartheta}\: \bfe_y+
    	\sin\vartheta\sin\varphi\:\bfe_z)\cdot\bfe_{\rm s}\\
    	(\cos\vartheta\: \bfe_x
    	-\sin\vartheta\cos\varphi\:\bfe_z)\cdot\bfe_{\rm s}
    	\end{array}\right],
	\label{hs=}
	\end{align}
where $(k,\vartheta_0,\varphi_0)$ and $(k,\vartheta,\varphi)$ are respectively the spherical coordinates of the incident and scattered wavevectors, $\bk_{\rm i}$ and $\bk_{\rm s}$. Notice that because we consider the scattering of a left-incident wave, $\vartheta_0\in[0,\frac{\pi}{2})$, $\bB_+(\vec p)=\bzero$, 
and $\bA_-(\vec p)=4\pi^2\delta(\vec p-\vec k_{\rm i})\bup_{\rm i}$, where $\vec k_{\rm i}$ is the projection of $\bk_{\rm i}$ onto the $x$-$y$ plane. The last equation together with (\ref{project-pm})  and (\ref{proj-Ids}) imply $\bPi_1(\vec k_{\rm i})\bup_{\rm i}=\bup_{\rm i}$ and $\bPi_2(\vec k_{\rm i})\bup_{\rm i}=\bzero$.

\item Let $\bT^l_\pm\in\sF^4_k$ be the 4-component functions satisfying
	 \begin{align}
    	&\widehat\bPi_1\bT^l_+=\bT^l_+,
    	\quad\quad\widehat\bPi_2\bT^l_-=\bT^l_-,
     	\quad\quad\widehat\bPi_1\bT^l_-=\widehat\bPi_2\bT^l_+=\bzero,
    	\label{id-21}\\
	&\widehat\bPi_2\,\widehat\bM\,\bT_-^l=
	-4\pi^2\widehat\bPi_2\big(\widehat\bM-\bI\big)
	\bup_{\rm i} \delta_{\vec k_{\rm i}},
	\label{Eq-TmL}\\
	&\bT^l_+=\widehat\bPi_1\big(\,\widehat{\bM}-\widehat\bI\big)\big(\bT^l_-+4\pi^2
	\bup_{\rm i} \delta_{\vec k_{\rm i}}\big),
	\label{Eq-TpL}
	\end{align}
where $\delta_{\vec k_{\rm i}}$ stands for the Dirac delta function in two dimensions that is centered at $\vec k_{\rm i}$, i.e., $\delta_{\vec k_{\rm i}}(\vec p):=\delta(\vec p-{\vec k_{\rm i}})$. As noted in Ref.~\cite{jpa-2020a}, we can represent $\widehat\bM$ as a $4\times 4$ matrix whose entries are integral operators. Therefore, (\ref{Eq-TmL}) is a system of linear integral equations for the components of $\bT_-^l$.
% In view of (\ref{id-21}), we can introduce $\widehat{\bM}_{ij}:=\widehat\bPi_i\widehat{\bM}\,\widehat\bPi_j$ and express (\ref{Eq-TmL}) and (\ref{Eq-TpL}) in the form,
%	\begin{align} 
%	&\widehat{\bM}_{22}\bT^l_-=-4\pi^2\widehat{\bM}_{21}\bup_{\rm i} \delta_{\vec k_{\rm i}},
%	\label{TmL=}\\
%	&\bT^l_+=\widehat{\bM}_{12}\bT^l_-+4\pi^2\big(\widehat{\bM}_{11}-\widehat\bI\big)\bup_{\rm i} \delta_{\vec k_{\rm i}}.
%	\label{TpL=}
%	\end{align}
%Again (\ref{TmL=}) is an integral equation for $\bT^l_-$.

\item Determine the scattering amplitude and differential cross section using 
	\begin{align}
    	&f(\bk_{\rm i},\bk_{\rm s}) \bfe_{\rm s}=
    	-\frac{ik|\cos\vartheta|}{2\pi}\,
   	\bXi^T\bT^l_\pm(\vec k_{\rm s})
	~~~~~~{\rm for}~~~\pm\cos\vartheta>0,
	\label{f=2}\\[6pt]
    	&\sigma_d(\bk_{\rm i},\bk_{\rm s})=\frac{k^2\cos^2\vartheta\,	\bT^l_\pm(\vec k_{\rm s})^{\dagger}\,
    	\bT^l_\pm(\vec k_{\rm s})}{4\pi^2 (1+\cos^2\vartheta)}
    	~~~{\rm for}	~~~\pm\cos\vartheta>0,
    \label{cross-sec=1}
    \end{align}
where
	\be
	\bXi:=\left[\begin{array}{c}
	\bfe_x\\
	\bfe_y\\
	\sin\theta\sin\varphi\,\bfe_z\\
	-\sin\theta\cos\varphi\,\bfe_z\end{array}\right],
	\label{bXi-def}
	\ee
and $\dagger$ stands for the complex-conjugate of the transpose of a matrix \cite{jpa-2020a}. It is not difficult to see that the term $\bXi^T\bT^l_\pm(\vec k_{\rm s})$ entering (\ref{f=2}) admits the following more explicit expression.
	\be
	\bXi^T\bT^l_\pm(\vec k_{\rm s})=\bt_\pm^++[(\bt_\pm^-\times\hat\bfr)\cdot\bfe_z]\bfe_z ~~~\for~~~\pm\cos\vartheta>0,
	\label{Xi-Tpm=}
	\ee
where 
	\begin{align}
	&\bt_\pm^+:=t_{\pm 1}\,\bfe_x+t_{\pm 2}\,\bfe_y,
	&&\bt_\pm^-:=t_{\pm 3}\,\bfe_x+t_{\pm 4}\,\bfe_y,
	\label{tpms}
	\end{align}
$t_{\pm m}$ denote the components of $\bT^l_\pm(\vec k_{\rm s})$, so that $\bT^{l}_\pm(\vec k_{\rm s})^T=\left[t_{\pm 1}~~t_{\pm 2}~~t_{\pm 3}~~t_{\pm 4}\right]$,
%	\be
%	\bT^{l}_\pm(\vec k_{\rm i})^T=\left[t_{\pm 1}~~t_{\pm 2}~~t_{\pm 3}~~t_{\pm 4}\right],
%	\label{Tpm-vec}
%	\ee 
and we have employed (\ref{bXi-def}) -- (\ref{tpms}). Notice that because $\bfe_s$ is a unit vector and $|f(\bk_{\rm i},\bk_{\rm s})|=\sqrt{\sigma_d(\bk_{\rm i},\bk_{\rm s})}$, we can use $f(\bk_{\rm i},\bk_{\rm s}) \bfe_{\rm s}/\sqrt{\sigma_d(\bk_{\rm i},\bk_{\rm s})}$ to determine $\bfe_s$ and $f(\bk_{\rm i},\bk_{\rm s})$  up to a physically irrelevant phase factor.
    
    \end{enumerate}

The above procedure reduces the solution of the EM scattering problems for a general linear scattering medium to the determination of the fundamental transfer matrix $\widehat\bM$ and the solution of  (\ref{Eq-TmL}). We can easily obtain a series solution of this equation. To do this, we introduce the operators,
	\be
	\widehat\bN_j:=(-1)^j\widehat\bPi_j(\widehat\bI-\widehat\bM),~~~~~j\in\{1,2\},
	\label{N=}
	\ee
and use (\ref{Eq-TmL}) and (\ref{Eq-TpL}) to show that
	\begin{align}
	&\bT^l_-=4\pi^2\widehat\bN_2\,(\widehat\bI-\widehat\bN_2)^{-1}\bup_{\rm i} \delta_{\vec k_{\rm i}}=	4\pi^2\widehat\bN_2\sum_{\ell=0}^\infty\widehat\bN_2^{\ell} \bup_{\rm i} \delta_{\vec k_{\rm i}},
	\label{series-m}\\
	&\bT^l_+=4\pi^2\widehat\bN_1\,(\widehat\bI-\widehat\bN_2)^{-1}\bup_{\rm i} \delta_{\vec k_{\rm i}}=	4\pi^2\widehat \bN_1\sum_{\ell=0}^\infty\widehat\bN_2^{\ell} \bup_{\rm i} \delta_{\vec k_{\rm i}}.
	\label{series-p}
	\end{align}
Substituting these relations in (\ref{f=2}) we arrive at a series expansion for $f(\bk_{\rm i},\bk_{\rm s}) \bfe_{\rm s}$. 
  
If the source of the incident wave is located at $z=+\infty$, the above procedure applies except that the role of $\bT_\pm^l$ is played by another pair of functions $\bT_\pm^r\in\sF^4_k$ that satisfy (\ref{id-21}) -- (\ref{Eq-TpL}), (\ref{series-m}), and (\ref{series-p}) for a $\bup_{\rm i}$ that is associated with a right-incident wave. The latter is given by (\ref{incident}) and (\ref{hi=}) with $\vartheta_0\in(\frac{\pi}{2},\pi]$. Furthermore, because for a right-incident wave, $\bA_+(\vec p)=\bzero$ and $\bB_+(\vec p)=4\pi^2\delta(\vec p-\vec k_{\rm i})\bup_{\rm i}$, Eqs.~(\ref{project-pm})  and (\ref{proj-Ids}) imply $\bPi_1(\vec k_{\rm i})\bup_{\rm i}=\bzero$ and $\bPi_2(\vec k_{\rm i})\bup_{\rm i}=\bup_{\rm i}$. 
%We can also show that
%	\begin{align}
%	&\widehat{\bM}_{22}\bT^r_-=4\pi^2\big(\widehat\bI-\widehat{\bM}_{22}\big)\!\bup_{\rm i}
%    	\delta_{\vec k_{\rm i}},
%   	 \label{Tm=2r}\\
%    	&\bT^r_+=\widehat{\bM}_{12} \big(\bT^r_-+4\pi^2\bup_{\rm i}\delta_{\vec k_{\rm i}}\big).
%    	\label{Tp=2r}
%	\end{align}
%\wyt{$\widehat{\bfM}\widehat\fM$}\vspace{-24pt}

\section{Auxilary transfer matrix and its Dyson expansion} 

We have defined the fundamental transfer matrix as a linear operator $\widehat\bM:\sF_k^4\to\sF_k^4$ that maps $\bC_-$ to $\bC_+$. According to (\ref{asym-EM-2}) and (\ref{cbC-bC}), these are given by the coefficient functions $\bcA_\pm$ and $\bcB_\pm$ determining the Fourier transform $\tilde\bPhi(\vec p,z)$ of $\bPhi(\vec r,z)$ with respect to $\vec r$ for $z\notin [a_-,a_+]$. 

Let us recall that $\bPhi(\vec r,z)$ satisfies the time-dependent Schr\"odinger equation (\ref{TD-sch-eq}) where $z$ plays the role of time. Clearly, the Hamiltonian operator $\widehat\bH$ entering this equation depends on $z$. Making this dependence explicit and viewing $\bPhi(\cdot,z)$  as a function that at `time' $z$ assigns to each $\vec r\in\R^2$ a column vector belonging to $\C^{4\times 1}$, we can write (\ref{TD-sch-eq}) in the form
	\be
	i\partial_z\bPhi(\cdot,z)=\widehat{\bH}(z)\bPhi(\cdot,z).
	\label{sch-eq}
	\ee
Let $\widehat\cF$ denote the (two-dimensional) Fourier transformation of functions of $\vec r$. This is a linear operator $\widehat\cF$ that maps $\bPhi(\cdot,z)$ to $\tilde\bPhi(\cdot,z)$. Applying $\widehat\cF$ to both sides of (\ref{sch-eq}), we find
	\be
	i\partial_z\tilde\bPhi(\cdot,z)=\widehat{\tilde\bH}(z)\tilde\bPhi(\cdot,z),
	\label{sch-eq-tilde}
	\ee
where $\widehat{\tilde\bH}(z):=\widehat\cF\,\widehat{\bH}(z)\widehat\cF^{-1}$. Given an initial value $z_0$ of $z$, we can express the solutions of (\ref{sch-eq-tilde}) in the form,
	\be
	\tilde\bPhi(\cdot,z)=\widehat{\tilde\bU}(z,z_0)\tilde\bPhi(\cdot,z_0),
	\label{evolution-1}
	\ee
where $\widehat{\tilde\bU}(z,z_0)$ is the evolution operator for the Hamiltonian $\widehat{\tilde\bH}(z)$, i.e.,
	\bea
	\widehat{\tilde\bU}(z,z_0)&:=&\sT\exp\left[-i\int_{z_0}^z \widehat{\tilde\bH}(z') dz'\right]\nn\\
	&:=&\widehat\bI+\sum_{\ell=1}^\infty (-i)^\ell
        \int_{z_0}^z \!\!dz_\ell\int_{z_0}^{z_\ell} \!\!dz_{\ell-1}
        \cdots\int_{z_0}^{z_2} \!\!dz_1\,
        \widehat{\tilde\bH}(z_\ell)\widehat{\tilde\bH}(z_{\ell-1})\cdots\widehat{\tilde\bH}(z_1),
	\eea
and $\sT$ stands for the `time-ordering' operator \cite{weinberg-qm}.

For $z\leq a_-$ and $z\geq a_+$, we have $\widehat{\bH}(z)=\widehat{\bH}_0$, (\ref{sch-eq-tilde}) reduces to (\ref{sch-eq-zero-FT}), and $\tilde\bPhi(\cdot,z)$ satisfies (\ref{asym-EM-2}). Performing the Fourier transform of both sides of this equation and using (\ref{eg-va}) and (\ref{cbC-bC}), we obtain		\be
	\tilde\bPhi(\cdot,z)= \left\{\begin{array}{ccc}
	e^{-iz\widehat{\tilde\bH}_0}\,\bcC_- &\for& z<a_-,\\
	e^{-iz\widehat{\tilde\bH}_0}\,\bcC_+ &\for& z>a_+.\end{array}\right.
	\label{P-C}
	\ee
Consequently,
	\begin{align}
	&\bcC_\pm=e^{ia_\pm\widehat{\tilde\bH}_0}\tilde\bPhi(\cdot,a_\pm).
	\label{Cs=}
	\end{align} 
In particular, introducing
	\be
	\widehat\bcU(z,z_0):=e^{iz\widehat{\tilde\bH}_0}\,\widehat{\tilde\bU}(z,z_0)\,
	e^{-iz_0\widehat{\tilde\bH}_0},
	\ee 
and using (\ref{evolution-1}) and (\ref{Cs=}), we have
	\be
	\bcC_+=\widehat\bcU(a_+,a_-)\bcC_-.
	\ee
Comparing this equation with (\ref{M-def}) and recalling that according to (\ref{cbC-bC}), $\bC_\pm=\widehat\Lpi\bcC_\pm$, we arrive at 
	\be
	\widehat\bM=\widehat\Lpi\,\widehat\bfM\,\widehat\Lpi, %\wyt{\widehat{\fM}}
	\label{M=PMP}
	\ee
where $\widehat{\bfM}:=\widehat\bcU(a_+,a_-)$ is the EM analog of the auxiliary transfer matrix of Ref.~\cite{pra-2021}.

It is easy to show that $\widehat\bcU(z,z_0)$ is the evolution operator for the (interaction-picture) Hamiltonian:
	\be
	\widehat\bcH(z):=e^{iz\widehat{\tilde{\bH}}_0}\left[\widehat{\tilde{\bH}}(z)-	
	\widehat{\tilde{\bH}}_0\right]e^{-iz\widehat{\tilde{\bH}}_0}.
    	\label{int-pic-H}
    	\ee
In other words, $\widehat\bcU(z,z_0)=\sT\exp\big[-i\int_{z_0}^z \widehat\bcH(z')dz'\big]$. If for some $z\in\R$, $\hat\bvarepsilon{\mbox{\normalsize$(\vec r,z)$}}=\hat\bmu{\mbox{\normalsize$(\vec r,z)$}}=\bI$ for all $\vec r\in\R^2$, then $\widehat{\bH}(z)=\widehat\bH_0$, $\widehat{\tilde\bH}(z)=\widehat{\tilde\bH}_0$, and (\ref{int-pic-H}) implies $\widehat\bcH(z)=\widehat\bzero$. Now, suppose that there is some interval $\cI\subseteq\R$ such that for all $z\in\cI$ and $\vec r\in\R^2$, $\hat\bvarepsilon{\mbox{\normalsize$(\vec r,z)$}}=\hat\bmu{\mbox{\normalsize$(\vec r,z)$}}=\bI$ . Then $\widehat\bcH(z)=\widehat\bzero$ for all $z\in\cI$, which in turn implies $\widehat\bcU(z_+,z_-)=\widehat\bI$ for all  $z_\pm\in \cI$. In particular, $\widehat\bcU(a_-,-\infty)=\widehat\bcU(+\infty,a_+)=\widehat\bI$,  $\widehat\bcU(+\infty,-\infty)=\widehat\bcU(a_+,a_-)$, and
	\be
        \mbox{$\widehat{\bfM}$}=\widehat\bcU(+\infty,-\infty)=\widehat\bI+\sum_{\ell=1}^\infty (-i)^\ell
        \int_{-\infty}^\infty \!\!dz_\ell\int_{-\infty}^{z_\ell} \!\!dz_{\ell-1}
        \cdots\int_{-\infty}^{z_2} \!\!dz_1\,
        \widehat{\bcH}(z_\ell)\widehat{\bcH}(z_{\ell-1})\cdots\widehat{\bcH}(z_1).
        \label{dyson}
	\ee
Substituting this equation in (\ref{M=PMP}) we find a series expansion for the fundamental transfer matrix $\widehat\bM$. This offers a method for computing $\widehat\bM$ which is particularly effective, if the Dyson series (\ref{dyson}) terminates. 
	
Another consequence of the vanishing of $\widehat\bcH(z)$ in empty space is the composition rule for the auxiliary transfer matrix $\widehat\bfM$; let $a_0,a_1,a_2,\cdots,a_n$ be an increasing sequence of real numbers such that $a_0=a_-$ and $a_{n}=a_+$, for all $l\in\{1,2,\cdots,n\}$, $\cI_l:=[a_{l-1},a_l]$, and $\widehat\bfM_l:=\widehat\bcU(a_l,a_{l-1})$ be the auxiliary transfer matrix for $\cI_l$, then
	\be
	\widehat\bfM_n\widehat\bfM_{n-1}\cdots \widehat\bfM_1=\widehat\bcU(a_n,a_{n-1})\widehat\bcU(a_{n-1},a_{n-2})\cdots \widehat\bcU(a_1,a_0)=\widehat\bcU(a_+,a_-)=\widehat\bfM.
	\label{compose}
	\ee
This is the EM analog of the composition property of the auxiliary transfer matrix for scalar waves \cite{pra-2021}.\footnote{Ref.~\cite{jpa-2020a} assumes that the contribution of the evanescent waves to the solution of the scattering problem is negligible. If this assumption holds, we can identify the fundamental and auxiliary transfer matrices, and (\ref{compose}) coincides with Eq.~(96) of Ref.~\cite{jpa-2020a}. Note however that this provides an approximate description of the scattering phenomenon which is exact for certain setups \cite{p158}. The principal example is the isotropic EM point scatterer and invisible configurations studied in Ref.~\cite{jpa-2020a}.}

\section{Scattering by a planar collection of point scatterers}

Consider a collection of $N$ non-magnetic point scatterers that lie on the $x$-$y$ plane and whose relative permittivity and permeability tensors have the form,
	\begin{align}
	&\hat\bvarepsilon(\vec r,z)=\bI+\delta(z)\sum_{a=1}^N\bfZ_a\,\delta(\vec r-\vec r_a),
	&&\hat\bmu\mbox{\normalsize$(\vec r,z)$}=\bI,
	\label{delta}
	\end{align}
where $\bfZ_a$ are $3\times 3$ complex matrices, and $\vec r_a=(x_a,y_a)$ signify the positions of the point scatterer in the $x$-$y$ plane, as shown in Fig.~\ref{fig1}.\footnote{Clearly, $\vec r_a=\vec r_b$ if and only if $a=b$.}

If $N=1$ and $\bfZ_1=\fz\,\bI$ for some $\fz\in\C$, (\ref{delta}) describes an isotropic point scatterer whose scattering problem has been studied thoroughly \cite{VCL}. The standard treatment of this problem requires dealing with certain divergent terms. This is usually achieved through a delicate regularization of these terms and a coupling-constant renormalization to subtract the unwanted singularities. In Ref.~\cite{jpa-2020a} we use the dynamical formulation of EM scattering to obtain an exact solution for this problem which avoids the singularities of the standard treatment and yields the same result. In this section, we use the results of Secs.~2 and 3 to explore the scattering properties of the collection of point scatterers corresponding to (\ref{delta}){\footnote{{Our approach also applies to more general (nonplanar) configurations of point scatterers, but for these configurations the exact and analytic treatment of the problem, i.e., the determination of the explicit form of the entries of the fundamental transfer matrix and the $4$-component functions $\bT^{l/r}_{\pm}$ that give the scattered wave becomes intractable.}}} for the generic case where $\bfZ_a$ are arbitrary $3\times 3$ complex matrices with a nonzero $\fZ_{a,33}$ entry;
	\be
	\fZ_{a,33}\neq 0~~~\mbox{for all}~~~a\in\{1,2,\cdots,N\}.
	\label{condi}
	\ee 
This is a technical condition that is extremely difficult to be violated by a physically realizable point scatterer.

\subsection{Calculation of the fundamental transfer matrix}

We begin our study of the point scatterers (\ref{delta}) by drawing attention to the following identities whose proof we give in Appendix~A.
	\bea
	&&\frac{\delta(\vec r-\vec r_a)}{1+\sum_{c=1}^N\fc_c\,\delta(\vec r-\vec r_c)}=0,
	\label{id-100a}\\[6pt]
	&&\frac{\fc_a\,\delta(\vec r-\vec r_a)\delta(\vec r-\vec r_b)}{1+\sum_{c=1}^N\fc_c\,\delta(\vec r-\vec r_c)}= \delta_{ab}\,\delta(\vec r-\vec r_a),
	\label{id-100b}
	\eea
where $a,b\in\{1,2,\cdots,N\}$, and $\fc_a$'s are nonzero complex numbers. In view of (\ref{new1}), (\ref{JEs=}), and (\ref{delta}) -- (\ref{id-100a}),
	 \begin{align}
	&{\widehat\epsilon_{33}}^{\ -1}-1 =0,
    	&&{\widehat\epsilon_{33}}^{\ -1}\vec\bigvarepsilon_3 ={\widehat\epsilon_{33}}^{\ -1}\vec J_\cE=\vec 0.
	\label{pt-e1}
    	\end{align}
It is important to realize that these relations hold in the sense of distributions \cite{strichartz}. In particular, although $\tilde\bJ_\cE:=\hat\varepsilon_{33}^{-1}\vec J_\cE\vec\bigvarepsilon_3^{\,T}$, the last equation in (\ref{pt-e1}) does not imply $\tilde\bJ_\cE=\bzero$. In fact, we can use (\ref{new1}), (\ref{JEs=}), (\ref{delta}), and (\ref{id-100b}) to show that
	 \begin{align}
    	\tilde\bJ_\cE
	%&=\delta(z)\sum_{a=1}^N 
	%\frac{\delta(\vec r-\vec r_a)}{\fZ_{a,33}}\left[\begin{array}{c}-\fZ_{a,23}\\
	%\fZ_{a,13}\end{array}\right]\left[\,\fZ_{a,31}~~\fZ_{a,32}\,\right]
	%\nn\\
	&=\delta(z)\sum_{a=1}^N 
	\frac{\delta(\vec r-\vec r_a)}{\fZ_{a,33}}\left[\begin{array}{cc}
	-\fZ_{a,23}\,\fZ_{a,31}&-\fZ_{a,23}\,\fZ_{a,32}\\
	\fZ_{a,13}\,\fZ_{a,31}&\fZ_{a,13}\,\fZ_{a,32}\,\end{array}\right],
	\label{pt-tJE}
    	\end{align}
where $\fZ_{a,ij}$ stand for the entries of $\bfZ_a$. Furthermore, according to (\ref{JEs=}) and (\ref{delta}),
	 \begin{align}
    	&\bJ_\cE=-i\bsigma_2+\delta(z)\sum_{a=1}^N 
	\delta(\vec r-\vec r_a)\!\left[\begin{array}{cc}
	-\fZ_{a,21} &-\fZ_{a,22} \\
	\fZ_{a,11} &\fZ_{a,12}\end{array}\right].
	\label{pt-JE}
    	\end{align}
	
Next, we use (\ref{new1}), (\ref{JHs=}), (\ref{tJs=}), and (\ref{delta}) to establish
	\begin{align}
	&{\widehat\mu_{33}}-1 =0,
    	&&{\widehat\mu_{33}}^{\ -1}\vec\bigmu_3 ={\widehat\mu_{33}}^{\ -1}\vec J_\cH=\vec 0,
	&& \bJ_\cH=-i\bsigma_2, && \tilde\bJ_{\cH}=\bzero.
	\label{pt-e2}
    	\end{align}
Substituting (\ref{pt-e1}) and (\ref{pt-e2}) in (\ref{H1}) and (\ref{H2}), and making use of (\ref{L-zero=}), we then find
	\begin{align}
	&\widehat\bH_{11}=\widehat\bH_{22}=\widehat\bzero,
	&&\widehat\bH_{12}=\widehat\bL_0,
	&&\widehat\bH_{21}=-\widehat\bL_0+k(\tilde\bJ_\cE-\bJ_\cE-i\bsigma_2).
	\label{pt-Hij}
	\end{align}
Eqs.~(\ref{H-def}), (\ref{H-zero}), (\ref{pt-tJE}), (\ref{pt-JE}), and (\ref{pt-Hij}) imply
	\be
	\widehat\bH-\widehat\bH_0=ik\delta(z)\,
	\sum_{a=1}^N\delta(\vec r-\vec r_a)
	\left[\begin{array}{cc}
	\bzero&\bzero \\
	\bZ_a\bsigma_2&\bzero\end{array}\right],
	\label{H-H0=}
	\ee
where
	\bea
	\bZ_a&:=&\frac{1}{\fZ_{a,33}}\left[\begin{array}{cc}
	\fZ_{a,22}\fZ_{a,33}-\fZ_{a,23}\fZ_{a,32}&\fZ_{a,23}\fZ_{a,31}-\fZ_{a,21}\fZ_{a,33}\\
	\fZ_{a,13}\fZ_{a,32}-\fZ_{a,12}\fZ_{a,33}&\fZ_{a,11}\fZ_{a,33}-\fZ_{a,13}\fZ_{a,31}
	\end{array}\right],
	\label{Z=}\\
	&=&\frac{1}{\fZ_{a,33}}\left[\begin{array}{cc}
	\cZ_{a,11}&-\cZ_{a,12}\\
	-\cZ_{a,21}&\cZ_{a,22}\end{array}\right],
	\nn
	\eea	
and $\cZ_{a,ij}$ stands for the minor of the $\fZ_{a,ij}$ entry of $\bfZ_a$,  i.e., $\cZ_{a,ij}$ is the determinant of the $2\times 2$ matrix obtained by deleting the $i$-th row and $j$-th column of $\bfZ_a$.

Next, we compute the Hamiltonian operator $\widehat\bcH(z)$ for our collection of point scatterer. In view of (\ref{int-pic-H}) and (\ref{H-H0=}),  
	\bea
	\widehat\bcH(z)&=&e^{iz\widehat{\tilde\bH}_0}\left[
	\cF(\widehat{\bH}-\widehat{\bH}_0)\cF^{-1}\right]e^{-iz\widehat{\tilde\bH}_0}\nn\\
	%&=&ike^{iz\widehat{\tilde\bH}_0}\delta(z)
	%\sum_{a=1}^N\tilde\delta(i\vec\nabla_p-\vec r_a)\left[\begin{array}{cc}
	%\bzero&\bzero \\
	%\bZ_a\bsigma_2&\bzero\end{array}\right]e^{-iz\widehat{\tilde\bH}_0},\nn\\
	&=&ik\delta(z)
	\sum_{a=1}^N\tilde\delta(i\vec\nabla_p-\vec r_a)\left[\begin{array}{cc}
	\bzero&\bzero \\
	\bZ_a\bsigma_2&\bzero\end{array}\right],
	\label{pt-sH=}
	\eea
where $\tilde\delta(i\vec\nabla_p-\vec r_a):\sF^m\to\sF^m$ is the linear operator given by
	\be
	\tilde\delta(i\vec\nabla_p-\vec r_a)\bF(\vec p)
	:=e^{-i\vec r_a\cdot\vec p}\,\widecheck\bF(\vec r_a),\nn
	\ee
$m$ is an arbitrary positive integer, $\bF\in\sF^m$, and $\widecheck\bF$ stands for the inverse Fourier transform of $\bF$, i.e., 
$\widecheck\bF(\vec r):=\frac{1}{4\pi^2}\int_{\R^2} d^2\vec p\,e^{i\vec r\cdot\vec p}\bF(\vec p)$.

A simple consequence of (\ref{pt-sH=}) is that for all $z_1,z_2\in\R$, $\widehat\bcH(z_2)\widehat\bcH(z_1)=\widehat\bzero$. Therefore, the Dyson series on the right-hand side of (\ref{dyson}) terminates and (\ref{M=PMP}) yields 
	\be
	\widehat\bM=\widehat\Lpi+
	k\sum_{a=1}^N\widehat\Lpi
	\tilde\delta(i\vec\nabla_p-\vec r_a)\widehat\Lpi\left[\begin{array}{cc}
	\bzero&\bzero \\
	\bZ_a\bsigma_2&\bzero\end{array}\right].
	\ee
If we view $\widehat\bM$ as a linear operator acting in $\sF^4_k$, we can write this relation in the form
	\be
	\widehat\bM=\widehat\bI+
	k\sum_{a=1}^N\widehat\Lpi
	\tilde\delta(i\vec\nabla_p-\vec r_a)\left[\begin{array}{cc}
	\bzero&\bzero \\
	\bZ_a\bsigma_2&\bzero\end{array}\right]=
	\left[\begin{array}{cc}
	\widehat\bI&\bzero \\
	k\sum_{a=1}^N\widehat\Lpi
	\tilde\delta(i\vec\nabla_p-\vec r_a)\bZ_a\bsigma_2&\widehat\bI\end{array}\right],
	\label{M-delta=}
	\ee
where 
	\begin{align}
	&\widehat\Lpi\tilde\delta(i\vec\nabla_p-\vec r_a)\bF(\vec p)=
	\chi_k(\vec p)e^{-i\vec r_a\cdot\vec p}\,\widecheck\bF(\vec r_a),
	&&\chi_k(\vec p):=\left\{\begin{array}{ccc}
	1&\for& |\vec p|<k,\\
	0&\for&|\vec p|\geq k.\end{array}\right.
	\label{tdelta}
	\end{align}
For the special case where all the point scatterers are isotropic, i.e., $\bfZ_a=\fz_a\,\bI$ for some $\fz_a\in\C$, (\ref{Z=}) gives $\bZ_a=\fz_a\bI$, and the transfer matrix is given by (\ref{M-delta=}) with $\bZ_a$ changed to $\fz_a$.

\subsection{Determination of the scattering amplitude and cross section}

The planar collection of point scatterers specified by (\ref{delta}) is clearly invariant under the reflection about the $x$-$y$ plane. This implies that the expression for the scattering amplitude for the left- and right-incident waves coincide. We therefore confine our attention to the scattering of a left-incident wave. 

First, we determine the four-component function $\bT_-^l$.

Because $\widehat\bPi_2\bT^l_-=\bT^l_-$, we can write (\ref{Eq-TmL}) in the form
	\be
	\bT^l_-= -\widehat\bPi_2(\widehat\bM-\bI)\left(\bT_-^l+4\pi^2\bup_{\rm i} \delta_{\vec k_{\rm i}}\right).
	\label{int-eq-m}
	\ee
Let $\vec T^\pm_-\in\sF^2_k$ be such that
	\be
	\bT^l_-=\left[\begin{array}{c}
	\vec T^+_-\\
	\vec T^-_-\end{array}\right].
	\label{T-2-comp}
	\ee
Then we can use (\ref{proj}), (\ref{incident}),  (\ref{M-delta=}), (\ref{tdelta}),  and (\ref{T-2-comp}) to show that, for all $\vec p\in\R^2$,
	\begin{align}
	&\big((\widehat\bM-\bI)(\bT_-^l+4\pi^2\bup_{\rm i} \delta_{\vec k_{\rm i}})\big)(\vec p)=k\left[\begin{array}{c}
	\vec 0\\
	\vec X(\vec p)\end{array}\right],
	\label{delta-93}\\
	&\vec T^+_-(\vec p)= -\frac{k}{2\varpi(\vec p)}\tilde\bL_0(\vec p)\vec X(\vec p),\quad\quad\quad
	\vec T^-_-(\vec p)= -\frac{k}{2}\vec X(\vec p),
	\label{Tpm}
	\end{align}	
where
	\begin{align}
	&\vec X(\vec p):=\chi_k(\vec p)\sum_{a=1}^N e^{-i\vec r_a\cdot\vec p}\,\bZ_a\bsigma_2\left(\vec x_a+ e^{i\vec k_{\rm i}\cdot\vec r_a}\vec\bfe_{\rm i}\right),
	\label{Xa=}
	\\
	&\vec x_a:=\widecheck{\vec T}^+_-(\vec r_a)=\frac{1}{4\pi^2}\int_{\R^2}d^2\vec p\,e^{i\vec r_a\cdot\vec p}\:\vec T^+_-(\vec p).
	\label{ta=}
	\end{align}
If we insert (\ref{Xa=}) in the first equation in (\ref{Tpm}) and use the result to evaluate the right-hand side of (\ref{ta=}), we find the following system of linear equations for $\vec x_a$. 
	\be
	\sum_{b=1}^N \bA_{ab}\,\vec x_b=\vec \fb_a, 
	\label{system}
	\ee
where 
	\begin{align}
	\bA_{ab}&:=\delta_{ab}\bI+ \bcL(\vec r_a-\vec r_b)\bZ_b\bsigma_2,
	\quad\quad\quad\quad\vec\fb_a:=- \sum_{b=1}^N e^{i\vec k_{\rm i}\cdot\vec r_b}
	\bcL(\vec r_a-\vec r_b)\bZ_b\bsigma_2\vec\bfe_{\rm i},
	\label{Aab-def}\\
	\bcL(\vec r)&:=\frac{k}{8\pi^2}\int_{\sD_k}d^2\vec p\:
	\frac{e^{i\vec r\cdot\vec p}\tilde\bL_0(\vec p)}{\varpi(\vec p)}
	=\frac{k}{8\pi^2}\widehat\bL_0
	\int_{\sD_k}d^2\vec p\:\frac{e^{i\vec r\cdot\vec p}}{\varpi(\vec p)}
	=\frac{k^2}{4\pi}\widehat\bL_0\,\sinc(k|\vec r|)\nn\\
	&=\frac{k}{4\pi}\left[\begin{array}{cc}
	\partial_x\partial_y&-\partial_x^2-k^2\\
	\partial_y^2+k^2 & -\partial_x\partial_y\end{array}\right]\sinc(k\sqrt{x^2+y^2}),
	\label{cL-def}
	\end{align}
and
	\begin{align}
	\sinc(x)&:=\sum_{n=0}^\infty \frac{(-1)^n x^{2n}}{(2n+1)!}=
	\left\{\begin{array}{ccc}
	\frac{\sin x}{x}&\for&x\neq 0,\\
	1&\for&x=0.\end{array}\right.
	\label{sinc-def}
	\end{align}
	
The system of equations (\ref{system}) has a unique solution, if there are matrices $\bB_{ab}\in\C^{2\times 2}$ satisfying
	\be
	\sum_{b=1}^N\bB_{ab}\bA_{bc}=\delta_{ac}\bI,
	\label{Bab-def}
	\ee
for all $a,c\in\{1,2,\cdots,N\}$. In this case, the solution of (\ref{system}) takes the form, $\vec x_a=\sum_{b=1}^N\bB_{ab}\vec\fb_b$, and (\ref{Xa=}) gives
	\be
	\vec X(\vec p)=\bsigma_2\,\vec{\bg}(\vec p),
	\label{sum-Xa}
	\ee
where
	\begin{align}
	&\vec{\bg}(\vec p):=\chi_k(\vec p)\sum_{a=1}^N e^{-i\vec r_a\cdot(\vec p-\vec k_{\rm i})}\bsigma_2\bZ_a\bsigma_2(\bI-\bx_a)\vec\bfe_{\rm i},
	\label{vec-g-def}\\
	&\bx_a:=\sum_{b,c=1}^N e^{-i\vec k_{\rm i}\cdot(\vec r_a-\vec r_c)}\bB_{ab}\bcL(\vec r_b-\vec r_c)\bZ_c\bsigma_2.
	\label{bx-a-def}
	\end{align}
In view of (\ref{proj}), (\ref{T-2-comp}), (\ref{Tpm}), and (\ref{sum-Xa}),
	\be
	\bT^l_-(\vec p)=-\frac{k}{2\varpi(\vec p)}\left[\begin{array}{c}
	\tilde\bL_0(\vec p)\bsigma_2\vec{\bg}(\vec p)\\
	\varpi(\vec p)\bsigma_2\vec{\bg}(\vec p)\end{array}\right].
	%=-k\,\widehat\bPi_2(\vec p)\bK\bup(\vec p).
	\label{Tm-delta}
	\ee  
Having obtained $\vec X(\vec p)$, we can calculate the right-hand side of (\ref{delta-93}) and use this equation together with (\ref{Eq-TpL}) to infer 
	\be
	\bT^l_+(\vec p)=
	\frac{k}{2\varpi(\vec p)}\left[\begin{array}{c}
	-\tilde\bL_0(\vec p)\bsigma_2\vec{\bg}(\vec p)\\
	\varpi(\vec p)\bsigma_2\vec{\bg}(\vec p)\end{array}\right].
	\label{Tp-delta}
	\ee 

Next, we introduce 
	\be
	\bJ:=\frac{i}{k}\tilde\bL_0(\vec k_{\rm s})\bsigma_2=
	\left[\begin{array}{cc}
	1-\sin^2\vartheta\cos^2\varphi & 
	-\sin^2\vartheta\sin\varphi\cos\varphi\\
    	-\sin^2\vartheta\sin\varphi\cos\varphi & 
	1-\sin^2\vartheta\sin^2\varphi
    	\end{array}\right],
	\label{J=2}
	\ee
and use (\ref{Tm-delta}) and (\ref{Tp-delta}) to etasblish 
	\be
	\bT^l_\pm(\vec k_{\rm s})=\frac{k}{2\varpi(\vec k_{\rm s})}\left[\begin{array}{c}
	ik\bJ\, \vec{\bg}(\vec k_{\rm s})\\
	\pm\varpi(\vec k_{\rm s})\bsigma_2\,\vec{\bg}(\vec k_{\rm s})\end{array}\right].
	\ee
This allows us to read off the components $t_{\pm m}$ of $\bT^l_\pm(\vec k_{\rm s})$ and determine the vectors $\bt^\pm_\pm$ of Eq.~(\ref{tpms}). With the help of (\ref{J=2}), we can write the result of this calculation in the form,
	\begin{align}
	&\bt^+_\pm=\frac{ik^2\left[\bg-(\hat\bfr\cdot\bg)\hat\bfr+
	(\hat\bfr\cdot\bg)(\bfe_z\cdot\hat\bfr)\bfe_z\right]}{2\varpi(\vec k_{\rm s})},
	&&\bt^-_\pm=\pm\frac{ik}{2}\,\bfe_z\times\bg,
	\label{tppm}
	\end{align}
where ${\bg}:=g_1\,\bfe_x+g_2\,\bfe_y,$ and $g_1$ and $g_2$ are the components of $\vec{\bg}(\vec k_{\rm s})$, so that $\vec{\bg}(\vec k_{\rm s})^T=[g_1~~g_2]$. Clearly, we can express $\bg$ as
	\be
	{\bg}= 
	\vec{\bg}(\vec k_{\rm s})^T\!
	\left[\begin{array}{c}
	1\\
	0\end{array}\right] \bfe_x+
	\vec{\bg}(\vec k_{\rm s})^T
	\!\left[\begin{array}{c}
	0\\
	1\end{array}\right]\bfe_y.
	\label{bg-def}
	\ee

Substituting (\ref{tppm}) in (\ref{Xi-Tpm=}), using the resulting expression in (\ref{f=2}), and noting that $\varpi(\vec k_{\rm s})=k|\cos\vartheta|$, $\bg\cdot\bfe_z=0$, and $\sigma_d(\bk_{\rm i},\bk_{\rm s})=|f(\bk_{\rm i},\bk_{\rm s}) \bfe_{\rm s}|^2$, we obtain
	\begin{align}
	&f(\bk_{\rm i},\bk_{\rm s}) \bfe_{\rm s}=
    	\frac{k^2}{4\pi}\,\big[\hat\bfr\times(\bg\times\hat\bfr)\big],
	\label{f=delta}\\
	&\sigma_d(\bk_{\rm i},\bk_{\rm s})=\frac{k^4}{16\pi^2}\left[|\bg|^2-|\hat\bfr\cdot\bg|^2\right].
	\label{sigma-delta}
	\end{align}
According to these relations, the information about the scattering properties of the collection of point scatterers described by (\ref{delta}) is contained in $\bg$. 

If the point scatterers are isotropic, so that $\bfZ_a=\fz_a\bI$ for some $\fz_a\in\C$, we have $\bZ_a=\fz_a\bI$, and Eqs.~(\ref{vec-g-def}) and (\ref{bx-a-def}) imply
	\be
	\vec\bg(\vec k_{\rm s})=\left[\sum_{a=1}^N 
	\fz_a e^{i(\vec k_{\rm i}-\vec k_{\rm s})\cdot\vec r_a}
	\bI-\sum_{a,b,c=1}^N \fz_c e^{i(\vec k_{\rm i}\cdot\vec r_c-
	\vec k_{\rm s}\cdot\vec r_a)}
	\bB_{ab}\bcL(\vec r_b-\vec r_c)\bsigma_2\right]\vec\bfe_{\rm i}.
	\label{vec-bg=-doublet}
	\ee
For a single isotropic point scatterer located at the origin of the coordinate system, i.e., $N=1$ and $\vec r_1=\vec 0$, we can use (\ref{Aab-def}), (\ref{cL-def}), (\ref{Bab-def}), and (\ref{vec-bg=-doublet}) to establish,
	\begin{align}
	&\bcL(\vec 0)=-\frac{ik^3}{6\pi}\,\bsigma_2,
	&&\bB_{11}=\bI-\bx_1=\bA_{11}^{-1}=\beta(\fz_1)\bI,
	&&\bg=\fz_1\beta(\fz_1)\vec e_{\rm i},
	\label{iso-delta}
	\end{align}
where
	\be
	\beta(\fz_1):=\left(1-\frac{i\fz_1 k^3}{6\pi}\right)^{\!\!-1}.
	\label{beta-def}
	\ee
{In particular, $\bg$ is a scalar multiple of the projection of the incident polarization vector onto the $x$-$y$ plane.} Plugging the last of Eqs.~(\ref{iso-delta}) in (\ref{f=delta}) and (\ref{sigma-delta}) we recover Eqs.~(118) and (119) of Ref.~\cite{jpa-2020a}.

For a single anisotropic point scatterer which is located at the origin and has its principal axes aligned with the $x$-, $y$- and $z$-axes, $N=1$, $\vec r_1=\vec 0$, and $\bfZ_1$ and $\bZ_1$ are diagonal matrices; there are $\fz_1,\fz_2,\fz_3\in\C$ such that 	
	\begin{align}
	&\bfZ_1=\left[\begin{array}{ccc}
	\fz_1 & 0 & 0\\
	0 & \fz_2 & 0\\
	0 & 0 &\fz_3\end{array}\right],
	&&\bZ_1=	\left[\begin{array}{cc}
	\fz_2 & 0\\
	0 & \fz_1\end{array}\right].\nn
	\end{align}
In view of these relations and (\ref{Aab-def}), (\ref{Bab-def}),
(\ref{vec-g-def}), (\ref{bx-a-def}), (\ref{bg-def}), and the first equation in (\ref{iso-delta}),
	\begin{align}
	&\bB_{11}=\bI-\bx_1=\bA_{11}^{-1}=\left[\begin{array}{cc}
	\beta(\fz_1) & 0 \\
	0 & \beta(\fz_2) \end{array}\right],
	&&\bg=\fz_1\beta(\fz_1) e_{{\rm i}\, x}\, \bfe_x+
	\fz_2\beta(\fz_2) e_{{\rm i}\,y}\, \bfe_y,
	\label{aniso-delta}
	\end{align}
where $e_{{\rm i}\, x}:=\bfe_{\rm i}\cdot\bfe_x$ and $e_{{\rm i}\, y}:=\bfe_{\rm i}\cdot\bfe_y$ are respectively the $x$- and $y$-components of the polarization vector $\bfe_{\rm i}$ for the incident wave. Eqs.~(\ref{f=delta}), (\ref{beta-def}), and (\ref{aniso-delta}) give
	\be
	f(\bk_{\rm i},\bk_{\rm s}) \bfe_{\rm s}=
    	\frac{k^2}{4\pi}\,\left[
	\frac{e_{{\rm i}\, x}\,\hat\bfr\times(\bfe_x\times\hat\bfr)}{\fz_1^{-1}-\frac{ik^3}{6\pi}}+
	\frac{e_{{\rm i}\, y}\,\hat\bfr\times(\bfe_y\times\hat\bfr)}{\fz_2^{-1}-\frac{ik^3}{6\pi}}\right].
	\label{f-delta-aniso}
	\ee
As is manifest from this equation, if $\fz_j=-6\pi i/k^3$ for $j=1$ (respectively $j=2$) the scattering amplitude blows up for $e_{\rm i\,x}\neq 0$ (respectively $e_{\rm i\,y}\neq 0$). This marks the emergence of a spectral singularity \cite{prl-2009} which corresponds to the situation where the point scatterer starts amplifying the background noise and emitting coherent EM waves \cite{pra-2011a}. According to (\ref{f-delta-aniso}), the presence of anisotropy affects the polarization of the emitted wave along the direction $\hat\bfr$.

\subsection{Scattering by a doublet of isotropic point scatterers}

\begin{figure}
        \begin{center}
        \includegraphics[scale=.4]{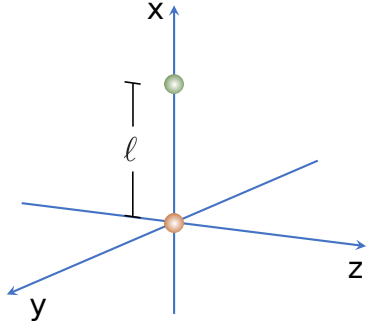}%\vspace{-5cm}
        \caption{Schematic view of a doublet of point scatterers lying on the $x$-axis.}
        \label{fig3}
        \end{center}
        \end{figure}

Consider a doublet of isotropic point scatterers that lie in the $x$-$y$ plane. We can always choose our coordinate system such that $\bfr_1=\bzero$ and $\bfr_2=\ell\,\bfe_x$ for a positive real parameter $\ell$, i.e., one of them is located at the origin and the other lies on the $x$ axis, as shows in Fig.~\ref{fig3}.
Let $\fz_1$ and $\fz_2$ be nonzero complex numbers such that $\bfZ_j=\fz_j\,\bI$  for $j\in\{1,2\}$. Then (\ref{Z=}) and (\ref{cL-def}) give 
	\begin{align}
	&\bZ_j=\fz_j\,\bI,
	&&\bcL(\vec 0)\bsigma_2=-\frac{ik^3}{6\pi}\bI,
	&&\bcL(\pm\bfr_2)\bsigma_2=-\frac{i}{4\pi\ell^3}\,\balpha,
	\label{cL-sigma2}
	\end{align}
where
	\begin{align}
	&\balpha:=\left[\begin{array}{cc}
	\alpha_1(k\ell) & 0\\
	0 & \alpha_2(k\ell)\end{array}\right],
	\label{alpha=}
	\end{align}\vspace{-18pt}
	\begin{align}
	&\alpha_1(x):=2(\sin x-x\cos x),
	&&\alpha_2(x):=(x^2-1)\sin x+x\cos x.
	\label{alphas=}
	\end{align}
Substituting (\ref{cL-sigma2}) in (\ref{Aab-def}), we have
	\begin{align}
	&\bA_{11}=\beta(\fz_1)^{-1}\bI,
	\quad\quad \bA_{12}=-\frac{i\fz_2}{4\pi\ell^3}\,\balpha,
	&&\bA_{21}=-\frac{i\fz_1}{4\pi\ell^3}\,\balpha,
	\quad\quad \bA_{22}=\beta(\fz_2)^{-1}\bI,
	\label{Aab-doublet}
	\\&\vec\fb_1=\frac{i}{12\pi\ell^3}\left[2(k\ell)^3\fz_1\bI+
	3 \fz_2 e^{ik_{{\rm i}x}\ell}\balpha\right]\vec\bfe_{\rm i},
	&&\vec\fb_2=\frac{i}{12\pi\ell^3}
	\left[2 (k\ell)^3\fz_2 e^{ik_{{\rm i}x}\ell}\bI+
	3\fz_1\balpha\right]\vec\bfe_{\rm i},
	\end{align}
where $k_{{\rm i}x}:=\bk_{\rm i}\cdot\bfe_x$. 

Because $\bA_{ab}$ are diagonal matrices, the determination of $\bB_{ab}$ is not difficult. In Appendix~B, we compute them for generic possibly anisotropic doublets of point scatterers. For the isotropic doublet we consider here, they take the form,
 	\be
	\begin{aligned}
	&\bB_{11}=\beta(\fz_1)\bgamma, 
	&&\bB_{12}=\frac{i\fz_2\beta(\fz_1)\beta(\fz_2)}{4\pi\ell^3}\,\balpha\bgamma,\\
	&\bB_{21}=\frac{i\fz_1\beta(\fz_1)\beta(\fz_2)}{4\pi\ell^3}\,\balpha\bgamma,
	\quad\quad\quad
	&&\bB_{22}=\beta(\fz_2)\bgamma, 
	\end{aligned}
	\label{Bab-doublet}
	\ee
where
	\begin{align}
	&\bgamma:=\left[\begin{array}{cc}
	\gamma_1 & 0\\
	0 & \gamma_2 \end{array}\right],
	&&\gamma_j:=\left[1+\frac{\fz_1\fz_2\,\beta(\fz_1)\beta(\fz_2)\alpha_j(k\ell)^2}{16\pi^2\ell^6}\right]^{-1}.
	\label{gamma-def}
	\end{align}
	
Next, we set $N=2$ in (\ref{vec-bg=-doublet}) and use (\ref{beta-def}), (\ref{cL-sigma2}), and (\ref{Bab-doublet}) to show that 
	\be
	\vec\bg(\vec k_{\rm s})=\bcG\,\vec\bfe_{\rm i},
	\label{g=Ge}
	\ee
where $\bcG$ is the diagonal $2\times 2$ matrix defined by
	\begin{align}
	&\bcG:=\fz_1\Big[\bI-\bG(\fz_1,\fz_2)-e^{-i\eta_{\rm s}}\bG_0\Big]+
	\fz_2 e^{i\eta_{\rm i}}\Big\{-\bG_0+ e^{-i\eta_{\rm s}}\big[\bI-\bG(\fz_2,\fz_1)\big]\Big\},
	\label{G=}
	\end{align}
and 
	\begin{align}
	&\bG(\fz_1,\fz_2):=\frac{\beta(\fz_1)}{48\pi^2\ell^3}\left[3\ell^{-3}\fz_2\beta(\fz_2)\balpha^2-8i\pi (k\ell)^3\bI\right]\bgamma,\quad\quad\quad
	\bG_0:=\frac{-i\beta(\fz_1)\beta(\fz_2)}{4\pi\ell^3}\,\balpha\bgamma,
	\label{G12}\\
	&\eta_{\rm i}:=\vec r_2\cdot\vec k_{\rm i}=\ell\,\bfe_x\cdot\bk_{\rm i}=k\ell\sin\theta_0\cos\varphi_0,
	\quad\quad\quad
	\eta_{\rm s}:=\vec r_2\cdot\vec k_{\rm s}=\ell\,\bfe_x\cdot\bk_{\rm s}=k\ell\sin\theta\cos\varphi.
	\end{align}
Substituting (\ref{beta-def}), (\ref{alphas=}), and (\ref{gamma-def}) in (\ref{G12}), we find an explicit formula for the matrix $\bcG$. Using this formula and Eqs.~(\ref{bg-def}) -- (\ref{sigma-delta}) and (\ref{g=Ge}), we can compute the scattering amplitude and differential cross-section for the system. This completes our treatment of the scattering problem for the doublets of isotropic point scatterers. In the remainder of this section we discuss its application in the study of the spectral singularities \cite{prl-2009} of these systems.

Spectral singularities correspond to the real values of the wavenumber $k$ for which the scattering amplitude blows up. Therefore, according to (\ref{f=delta}) they correspond to singularities of $\bg$ or equivalently $\vec\bg(\vec k_{\rm s})$. Inspecting Eqs.~(\ref{beta-def}), (\ref{alphas=}), and (\ref{gamma-def}) -- (\ref{G12}), we observe that $\vec\bg(\vec k_{\rm s})$ develops a singularity only if $\gamma_j$ blows up for $j=1$ or $j=2$, while $\beta(\fz_1)$ and $\beta(\fz_2)$ take finite values.\footnote{When $\gamma_j$ blows up, there may still be exceptional values of the angles $\theta_0,\varphi_0,\theta$, and $\varphi$ and polarizations of the incident wave for which $\vec\bg(\vec k_{\rm s})$ vanishes.} By virtue of (\ref{beta-def}) and (\ref{gamma-def}), this is equivalent to the requirement that the condition,
	\be
	\alpha_j(k\ell)^2=%- \frac{16\pi^2\ell^6}{{\fz_1\fz_2\beta(\fz_1)\beta(\fz_2)}}=
	-16\pi^2\ell^6
	\left(\fz_1^{-1}-\frac{ik^3}{6\pi}\right)\left(\fz_2^{-1}-\frac{ik^3}{6\pi}\right)\neq 0,
	\label{SS}
	\ee
holds for $j=1$ or $j=2$. Because $\alpha_j(x)$ take real values, (\ref{SS}) implies that $(\fz_1^{-1}-\frac{ik^3}{6\pi})(\fz_2^{-1}-\frac{ik^3}{6\pi})$ must be a negative real number. Let $\rho_j:=\RE(\fz_j)$ and $\varsigma_j:=\IM(\fz_j)$, so that $\fz_j=\rho_j+i\varsigma_j$. Then the latter condition is equivalent to demanding that one of the following holds.
	\begin{align}
	&{\rm C}_1:~~~~\rho_1=\rho_2=0~~{\rm and}~~\varsigma_1\varsigma_2\left(1+\frac{k^3\varsigma_1}{6\pi}\right)
	\left(1+\frac{k^3\varsigma_2}{6\pi}\right)>0;\nn\\
	&{\rm C}_2:~~~~\rho_1\rho_2<0~~{\rm and}~~\frac{1}{|\rho_1|}\left[\varsigma_1+ \frac{k^3}{6\pi}(\rho_1^2+\varsigma_1^2)\right]=\frac{1}{|\rho_2|}\left[\varsigma_2+ \frac{k^3}{6\pi}(\rho_2^2+\varsigma_2^2)\right].
	~~~~~~~~~~~~~~~~~~~~~~~\nn
	\end{align}
Notice that these are necessary conditions for the realization of a spectral singularity. Once ${\rm C}_1$ or ${\rm C}_2$ holds, we should in addition enforce (\ref{SS}) for either $j=1$ or $j=2$. Unlike ${\rm C}_1$ and ${\rm C}_2$, (\ref{SS}) restricts the distance $\ell$ between the point scatterers. Moreover, because $\alpha_j(x)$ involve trigonometric functions, for fixed values $\ell$, $\rho_j$, and $\varsigma_j$, the values of $k$ that fulfill (\ref{SS}) form a discrete set.

If (\ref{SS}) holds for $j=1$ (respectively $j=2$),  $\gamma_1=\infty$ (respectively $\gamma_2=\infty$). For generic values of $\theta_0,\varphi_0,\theta$, and $\varphi$, this implies $\cG_{11}=\infty$ (respectively $\cG_{22}=\infty$), where $\cG_{ij}$ are the entries of $\bcG$. 
In view of (\ref{g=Ge}), this condition identifies a spectral singularity provided that $\bfe_{\rm i}$ has a nonzero $x$-components (respectively $y$-component), i.e., $e_{{\rm i}\,x}\neq 0$ (respectively $e_{{\rm i}\,y}\neq 0$). For applications in optics, the exceptional values of $\theta_0,\varphi_0,\theta$, $\varphi$, and $\bfe_{\rm i}$ for which (\ref{SS}) fails to ensure the emergence of a spectral singularity are of no interest, because the system amplifies the background noise whose wave vector and polarization take all possible values. 

Clearly, ${\rm C}_2$ is less restrictive than ${\rm C}_1$. Demanding that ${\rm C}_2$ holds, we can fix one of $\rho_1,\rho_2,\varsigma_1$, and $\varsigma_2$ in terms of the other three and $k$. This leaves us with a total of four free parameters and the distance $\ell$ which enters (\ref{SS}). If we solve this equation for $k$ we find a discrete set of values of $k$ each depending on $\ell$ and three of $\rho_1,\rho_2,\varsigma_1$, and $\varsigma_2$. Suppose for definiteness that we use ${\rm C}_2$ to fix $\varsigma_2$, which we can relate to the gain coefficient for the point scatterer located at $\vec\bfr_2$. Then we can use (\ref{SS}) to express $k$ in terms of $\rho_1,\rho_2,\varsigma_1,\ell$ and possibly a discrete label counting the solutions of (\ref{SS}).

Finally notice that ${\rm C}_2$ obstructs the existence of spectral singularities for doublets consisting of identical ($\fz_1=\fz_2$) and $\cP\cT$-symmetric ($\fz_1=\fz_2^\star$) pairs of point scatterers.\footnote{The scattering problem for scalar waves interacting with a $\cP\cT$-symmetric pairs of point scatterers in three dimensions has been considered in Ref.~\cite{RB}. A proper treatment of this problem that yields its exact solution is given in \cite{p172}. For a general discussion of the scattering of scalar waves by doublets of point scatterers in two and three dimensions, see \cite{ap-2022}.} This shows that they cannot function as a laser unless they satisfy ${\rm C}_1$. In particular, the real part of their permittivity must equal that of vacuum; $\RE[\bvarepsilon(\bfr)]=\varepsilon_0\bI$. Materials satisfying this condition cannot usually display large enough gains so that the imaginary part their permittivity profile can be modeled using a delta function \cite{silfvast}. In the next section we discuss a class of doublet systems which readily satisfy ${\rm C}_2$.

\section{Active anti-$\cP\cT$-symmetric pairs of point scatterers, and their lasing threshold and spectrum}

Consider an anti-$\cP\cT$-symmetric doublet of isotropic point scatterers, which by definition \cite{Ge-2013,Wu-2015} satisfies $\fz_2=-\fz_1^*$ or equivalnetly,
	\begin{align}
	&\rho_1=-\rho_2,
	&&\varsigma_1=\varsigma_2.
	\end{align} 
Then ${\rm C}_2$ holds, and (\ref{SS}) identifies the values of $k\ell$ corresponding to spectral singularities with the real and positive solutions of the transcendental equations:
	\begin{align}
	&(\sin x-x\cos x)^2= \frac{1}{9}\left(x^6+c\,d\, x^3+c\right) ,
	\label{eq128-1}\\
	&[(x^2-1)\sin x+x\cos x]^2= \frac{4}{9}\left(x^6+c\,d\, x^3+c\right) ,
	\label{eq128-2}
	\end{align}
where $c:=36\pi^2\ell^6/(\rho_1^2+\varsigma_1^2)$, $d:= \varsigma_1/3\pi\ell^3$, and we have made use of (\ref{alphas=}). 

\begin{figure}
        \begin{center}
        \includegraphics[scale=.60]{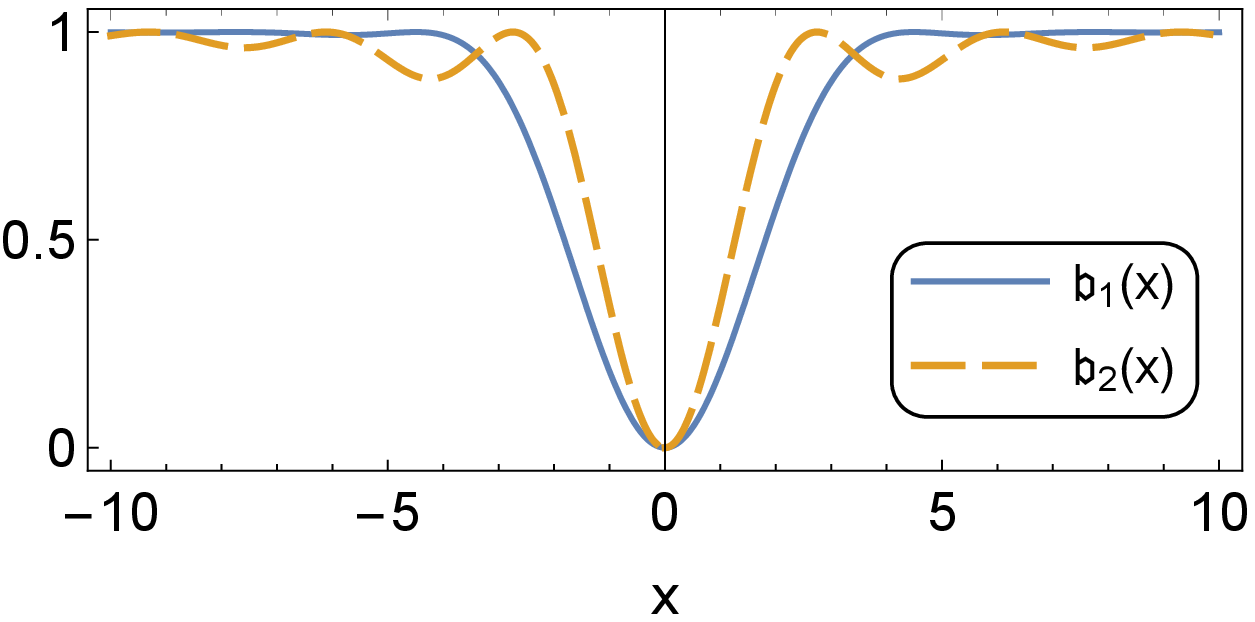}%\vspace{-5cm}
        \caption{Plots of $\fb_1$ and $\fb_2$.}
        \label{fig4}
        \end{center}
        \end{figure}

Let us introduce,
	\begin{align}
	&\fa_1(x):=\frac{\alpha_1(x)^2}{4}=(\sin x-x\cos x)^2,
	&&\fa_2(x):=\frac{\alpha_2(x)^2}{4}=\frac{[(x^2-1)\sin x+x\cos x]^2}{4}.
	\label{qs=}
	\end{align}
Then, we can write (\ref{eq128-1}) and (\ref{eq128-2}) in the form,\footnote{Eqs.~(\ref{eq128-1}) and (\ref{eq128-2}) correspond to (\ref{app-C1}) with $j=1$ and $j=2$, respectively.} 	 
	\be
	\fa_j(x)=\frac{1}{9}\left(x^6+\frac{12\pi \ell^3\varsigma_1 x^3}{\rho_1^2+\varsigma_1^2}+
	\frac{36\pi^2\ell^6}{\rho_1^2+\varsigma_1^2}\right).
	\label{app-C1}
	\ee
It is easy to see that this is a quadratic equation for $\varsigma_1$. We can express its solutions as
	\be
	\varsigma_1=-\frac{6\pi\ell^3}{\fb_j(x)x^3}\left[1\pm\sqrt{\Delta_j(x)}\right],
	\label{app-C2}
	\ee
where 
	\begin{align}
	&\fb_j(x):=1-\frac{9\,\fa_j(x)}{x^6},
	&&\Delta_j(x):=1-\fb_j(x)-\left(\frac{\rho_1^2x^6}{36\pi^2\ell^6}\right)\fb_j(x)^2.\nn
	\end{align}
The functions $\fb_j(x)$, whose graphs we plot  in Fig.\ref{fig4}, 
are continuous, vanish only at $x=0$, and satisfy
	\be
	0<\fb_j(x)\leq 1~~~\for~~~x\neq 0.
	\label{bound}
	\ee
	
According to (\ref{app-C2}) and (\ref{bound}), there are real and positive values of $x$ and $\ell$ and real values of $\rho_1$ and $\varsigma_1$ satisfying (\ref{app-C1}) only if $\Delta_j(x)\geq 0$. Suppose that this is the case. Then (\ref{bound}) implies $\Delta_j(x)\leq 1$. Using this in (\ref{app-C2}), we arrive at $\varsigma_1<0$. This argument shows that a spectral singularity can exist only for the negative values of the imaginary part of the permittivity of the point scatterers. This happens if the system has gain \cite{silfvast}. To produce and maintain this gain, the system must receive energy. We arrive at the same conclusion by noting that the presence of a spectral singularity leads to the emission of out-going EM waves which requires a source of energy. Therefore the mathematical result pertaining the nonexistence of the real and positive solutions of  (\ref{eq128-1}) and (\ref{eq128-2}) for $\varsigma_1\geq 0$ agrees with a physical requirement imposed by energy conservation.

\begin{figure}
        	\begin{center}
        	\includegraphics[scale=.6]{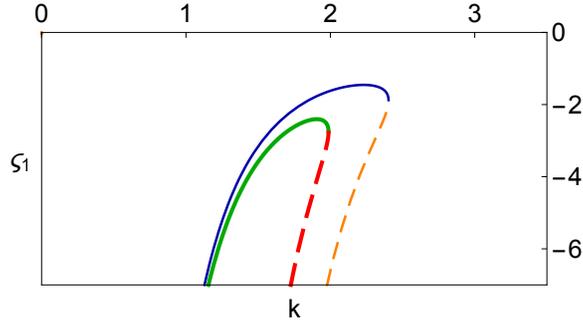}%\vspace{-5cm}
        	\caption{Location of spectral singularities in the $\varsigma_1$-$k$ plane for $\rho_1=1$ in units where $\ell=1$. The thin blue and dashed orange (thick green and dashed red) curves respectively correspond to setting $j=1$ ($j=2$) and choosing $-$ and $+$ signs on the right-hand side of (\ref{app-C2}). The critical values $\varsigma_+$ and $\varsigma_-$ of $\varsigma$ correspond to the maxima of the thin blue and thick green curves, respectively. The maximum value of $k$ for a spectral singularity is associated with the intersection of the thin blue and dashed orange curves where $\Delta_1=0$.}
        \label{fig5}
        \end{center}
        \end{figure}

To acquire more detailed information about the conditions ensuring the existence of spectral singularities for our anti-$\cP\cT$-symmetric doublet, we have used (\ref{app-C2}) to plot $\varsigma_1$ as a function of $k$ for fixed values of $\rho_1$ and $\ell$. Fig~\ref{fig5} shows the behavior of $\varsigma_1$ for $\rho_1=1$ in units where $\ell=1$.
It reveals the existence of a pair of critical values $\varsigma_{\pm}$ of $\varsigma_1$, with $\varsigma_{-}<\varsigma_{+}<0$, such that the following assertions hold. 
	\begin{itemize}
	\item[-] For $\varsigma_1>\varsigma_{+}$,  (\ref{eq128-1}) and (\ref{eq128-2}) have no real and positive solutions, i.e., no spectral singularities exists. 
	
	\item[-] For $\varsigma_1=\varsigma_{+}$ (\ref{eq128-1}) has a unique real and positive solution while (\ref{eq128-2}) has none. Therefore the system has a single spectral singularity. This establishes the existence of a threshold gain \cite{pra-2011a} and identifies, $\varsigma_1=\varsigma_{+}$, as the laser threshold condition for the system \cite{silfvast}.
		
	\item[-] For $\varsigma_{-}<\varsigma_1<\varsigma_{+}$, (\ref{eq128-1}) has two real and positive solutions, (\ref{eq128-2}) has no solutions, and the system develops two spectral singularities.
		
	\item[-] For $\varsigma_{1}=\varsigma_{-}$, (\ref{eq128-1}) and (\ref{eq128-2}) have respectively two and one real and positive solutions. So the number of spectral singularities of the system is three.
		
	\item[-] For $\varsigma_{1}<\varsigma_{-}$, (\ref{eq128-1}) and (\ref{eq128-2}) have two real and positive solutions each, and there are a total of four spectral singularities. 
	
	\item[-] The $k\ell$ values associated with spectral singularities have an upper bound. This is attained at a particular value of $\varsigma_1$ that lies between $\varsigma_{\pm}$. 
	
	\end{itemize}
For the particular example considered in Fig.~\ref{fig5}, $\varsigma_+=-1.454\ell^3$ and $\varsigma_-=-2.405\ell^3$. These identify spectral singularities with $k=2.230\ell^{-1}$ and $k=1.905\ell^{-1}$, respectively. The upper bound on the wavenumber signifying a spectral singularity is $2.403\ell^{-1}$. This is realized for $\varsigma_1=-1.884\ell^3$ which lies between $\varsigma_\pm$. Notice also that the spectral singularity obtained using larger values of $\varsigma_1$ have lower wavenumbers. 

\begin{figure}
        	\begin{center}
        	\includegraphics[scale=.50]{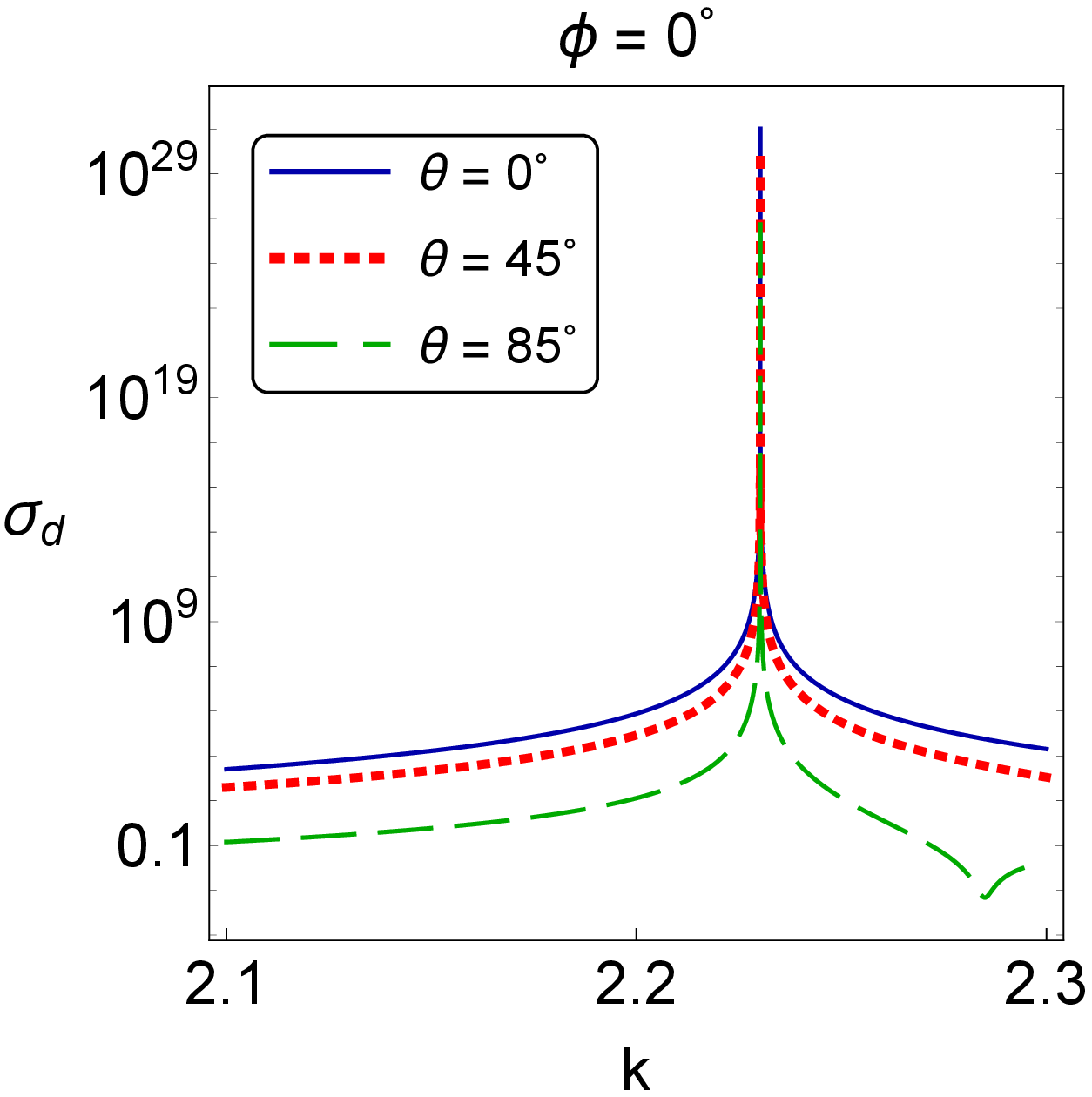}~~~~~~~
	\includegraphics[scale=.50]{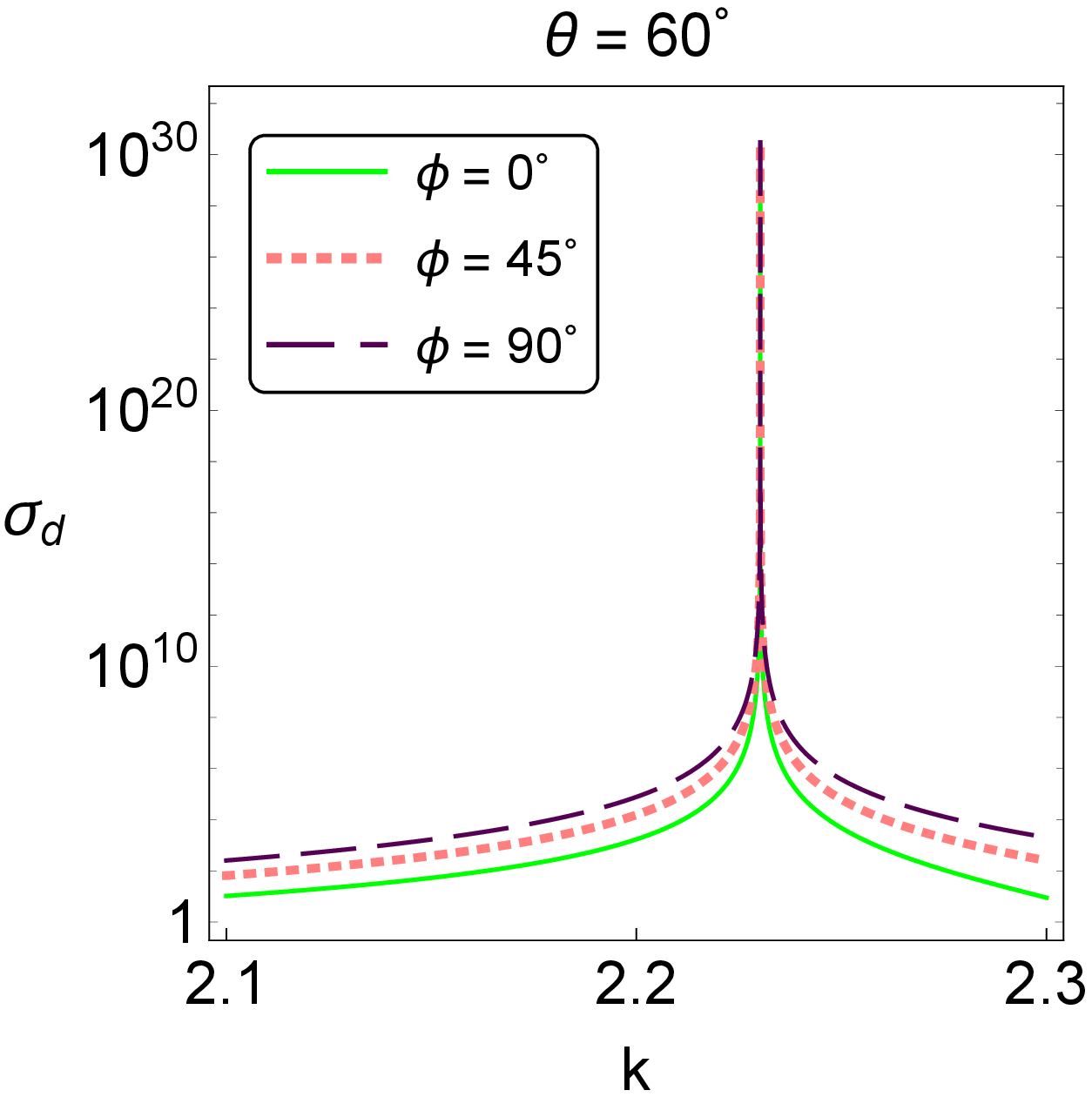}
        	\caption{Plots of the differential cross-section $\sigma_d(\bk_{\rm i},\bk_{\rm s})$ as a function of $k$ in units where $\ell=1$ for $\rho_1=1$, $\varsigma_1=-1.454$, $\hat\bk_{\rm i}=\bfe_z$  (i.e., $\vartheta_0=\varphi_0=0^\circ$), $\bfe_{\rm i}=\bfe_x$, and the different values for $\varphi$ and $\vartheta$.}
        \label{fig6}
        \end{center}
        \end{figure}

Fig.~\ref{fig6} provides graphical demonstrations of the spectral singularity corresponding to the lasing threshold, i.e., when $\rho_1=\ell^3$ and $\varsigma_1=\varsigma_+=-1.454\ell^3$. It shows the plots of the differential cross-section $\sigma_d(\bk_{\rm i},\bk_{\rm s})$ as a function of the wavenumber $k$ for $\hat\bk_{\rm i}=\bfe_z$ and different choices for $\bfe_{\rm i}$ and $\bk_{\rm s}$. The high peak represents the spectral singularity. It occurs for $k=2.230\ell^{-1}$ confirming our numerical calculation of this quantity which we have obtained by determining the maximum point of the blue curve shown in Fig.~\ref{fig5}.

An important property of the active anti-$\cP\cT$-symmetric doublets of point scatterers is that the wavenumber $k$ for their spectral singularities has a continuous dependence on $\varsigma_1$ except at $\varsigma_\pm$ where there is a jump on the number of spectral singularities. This shows that the system has a continuous lasing spectrum, a remarkable feature which was previously noticed in the study of certain one-dimensional anti-$\cP\cT$-symmetric optical systems \cite{Ge-2013,Hu-2021}.

\section{Concluding Remarks}

Scattering of EM waves by linear stationary media has been a focus of research activity since the late 19th century. The standard treatment of this subject involves the use of (dyadic) Green's functions \cite{TKD}. In the present article we have outlined an alternative approach that is based on a fundamental notion of transfer matrix. Unlike the traditional transfer matrices employed in the study of the EM wave propagation, this is not a numerical matrix, but a linear operator acting in an infinite-dimensional function space. This underlines the utility of the fundamental transfer matrix in performing analytic calculations. 

The approach we have presented here takes into account the contribution of the evanescent waves and applies to isotropic as well as anisotropic scatterers. Its central ingredient is the expression of the fundamental transfer matrix  in terms of the evolution operator for an effective non-unitary quantum system. In particular, we can expand it in a Dyson series. This makes it into an ideal tool for the study of scatterers where this series terminates. A physically important example is the class of point scatterers lying on a plane which by an appropriate choice of coordinates we take to be perpendicular to the scattering ($z$-) axis. For these systems we have computed the fundamental transfer matrix and used it to obtain an analytic closed-form expression for the scattering amplitude. The latter is given in terms of a vector $\bg$ that lies in the plane containing the point scatterers. A closer look at the structure of $\bg$ shows that its determination requires the inversion of a $2N\times 2N$ matrix (having $\bA_{ab}$ as its $2\times 2$ blocks) and performing a triple sums over the product of $2\times 2$ matrices carrying labels running over $\{1,2\cdots,N\}$. This suggests that it should be easy to use our analytic results to conduct reliable numerical studies of the scattering features of two-dimensional lattices and random ensembles of large number of point scatterers \cite{wang-2020}. 

A major disadvantage of the previous studies of EM point scatterers is the complications related to the emergence of singularities which required intricate regularization and renormalization schemes to deal with \cite{calla-2014,VCL}. Moreover, these studies only considered isotropic point scatterers. In contrast,  our treatment is free of unwanted singularities and applies to both isotropic as well as anisotropic point scatterers. Notice also that it does not make use of Foldy-Twersky's ansatz or the results of multiple scattering theory \cite{calla-2014,ishimaru2,martin}. 

We have conducted a detailed study of doublets consisting of a pair of isotropic point scatterers. In particular, we have addressed the problem of characterizing their spectral singularities. For the generic cases where the real part of the permittivity of the point scatterers is different from the vacuum permittivity, we have shown that doublets made of identical or $\cP\cT$-symmetric point scatterers do not admit a spectral singularity. This shows that they cannot function as a laser. This is not the case for anti-$\cP\cT$-symmetric doublets. We have provided a detailed analysis of the spectral singularities of the latter and determined their laser threshold condition and lasing spectrum. Similarly to the  one-dimensional anti-$\cP\cT$-symmetric slab systems studied in the literature \cite{Ge-2013,Hu-2021}, the anti-$\cP\cT$-symmetric doublet systems have a continuous lasing spectrum.

\section*{Acknowledgements}
We wish to express our gratitude to Alper Kiraz and Alphan Sennaro\u{g}lu for helpful discussions. This work has been supported by the Scientific and Technological Research Council of T\"urkiye (T\"UB\.{I}TAK) in the framework of the project 120F061 and by Turkish Academy of Sciences (T\"UBA).

\section*{Appendix A: Proof of (\ref{id-100a}) and (\ref{id-100b})}

Let $f:\R^2\to\C$ be any smooth function that has a compact support, and
	\begin{align}
	&\alpha_a(\vec r):=\frac{\delta(\vec  r-\vec r_a)}{1+\sum_{c=1}^N\fc_c\,\delta(\vec  r -\vec r_c)},
	&&\beta_{ab}(\vec r):=\frac{\fc_a\delta(\vec  r-\vec r_a)
	\delta(\vec  r-\vec r_b)}{1+\sum_{c=1}^N\fc_c\,\delta(\vec  r -\vec r_c)}.
	\label{alpha-beta}
	\end{align}
Then, because $\fc_a$'s do not vanish and $\vec r_a$'s are distinct elements of $\R^2$, 
	\bea
	\int_{\R^2}d^2\vec r^{\,\prime}\: f(\vec  r^{\,\prime})
	\alpha_a(\vec  r^{\,\prime}-\vec r)&=&
	\frac{f(\vec r-\vec r_a)}{1+\sum_{c=1}^N\fc_c\,\delta(\vec r_a-\vec r_c)}=\frac{f(\vec r-\vec r_a)}{1+\fc_a\,\delta(\vec 0)}
	=0,
	\label{id-100a-proof}\\[6pt]
	\int_{\R^2}d^2\vec r^{\,\prime}\: f(\vec  r^{\,\prime})
	\beta_{ab}(\vec  r^{\,\prime}-\vec r)&=&
	\frac{\fc_a\,f(\vec r-\vec r_a)\delta(\vec r_a-\vec r_b)}{\sum_{c=1}^N\fc_c\,\delta(\vec r_a-\vec r_c)}=\delta_{ab}f(\vec r-\vec r_a)\nn\\
	&=&\int_{\R^2}d^2\vec  r^{\,\prime}\: f(\vec  r^{\,\prime})\left[\delta_{ab}\delta(\vec  r^{\,\prime}-\vec r) \right].
	\label{id-100b-proof}
	\eea
Eqs.~(\ref{alpha-beta}) -- (\ref{id-100b-proof}) prove (\ref{id-100a}) and (\ref{id-100b}).

\section*{Appendix B: $\bB_{ab}$ for doublets of point scatterers} 

We can solve the system of equations (\ref{system}) using Gaussian elimination. In the following, we present the details of the solution for a doublet of point scatterers ($N=2)$ subject to the condition that $\bA_{aa}$ is invertible for both $a=1$ and $2$;
	\be
	\det\bA_{aa}\neq 0~~~\for~~~a\in\{1,2\}.
	\label{condi-det}
	\ee
	
When $N=2$, (\ref{system}) reads, 
	\bea
	\bA_{11}\vec x_1+\bA_{12}\vec x_2&=&\vec\fb_1,
	\label{app-b1}\\
	\bA_{21}\vec x_1+\bA_{22}\vec x_2&=&\vec\fb_2.
	\label{app-b2}
	\eea
Multiplying both sides of (\ref{app-b2}) by $-\bA_{12}\bA_{22}^{-1}$ and adding the resulting equation to (\ref{app-b1}), we find
	\be
	\big(\bA_{11}-\bA_{12}\bA_{22}^{-1}\bA_{21}\big)\vec x_1=
	\vec\fb_1-\bA_{12}\bA_{22}^{-1}\vec\fb_2.
	\label{app-b3}
	\ee
Similarly, multiplying both sides of (\ref{app-b1}) by $-\bA_{21}\bA_{11}^{-1}$ and adding the resulting equation to (\ref{app-b2}) give
	\be
	\big(\bA_{22}-\bA_{21}\bA_{11}^{-1}\bA_{12}\big)\vec x_2=
	\vec\fb_2-\bA_{21}\bA_{11}^{-1}\vec\fb_1.
	\label{app-b4}
	\ee
Eqs.~(\ref{app-b3}) and (\ref{app-b4}) have unique solutions provided that the matrices multiplying $\vec x_1$ and $\vec x_2$ are invertible. In this case the solution takes the form, $\vec x_a=\sum_{b=1}^2\bB_{ab}\vec\fb_b$, for
	\begin{align}
	&\bB_{11}=\big(\bA_{11}-\bA_{12}\bA_{22}^{-1}\bA_{21}\big)^{-1},
	&&\bB_{12}=-\bB_{11}\bA_{12}\bA_{22}^{-1},\\
	&\bB_{22}=\big(\bA_{22}-\bA_{21}\bA_{11}^{-1}\bA_{12}\big)^{-1},
	&&\bB_{21}=-\bB_{22}\bA_{21}\bA_{11}^{-1}.
	\end{align}

\ed